\documentclass[%
 aip,
 amsmath,amssymb,
 reprint,%
]{revtex4-1}

\usepackage{graphicx}
\usepackage{dcolumn}
\usepackage{bm}

\usepackage[utf8]{inputenc}
\usepackage[T1]{fontenc}
\usepackage{mathptmx}
\usepackage{etoolbox}

\makeatletter
\def\@email#1#2{%
 \endgroup
 \patchcmd{\titleblock@produce}
  {\frontmatter@RRAPformat}
  {\frontmatter@RRAPformat{\produce@RRAP{*#1\href{mailto:#2}{#2}}}\frontmatter@RRAPformat}
  {}{}
}%
\usepackage{color}
\usepackage{placeins}
\usepackage{tikz}
\usepackage{subfigure}
\usepackage{comment}
\usepackage{amsfonts}
\usepackage{multirow}
\usepackage{cancel}
\definecolor{nogo}{rgb}{1,0.2,0.3}
\definecolor{go}{rgb}{0,0.5,0.0}
\def \l{\lambda}
\def \hf{\text{HF}}
\def \ks{\text{KS}}

\def \sd{\text{SD}}
\def \ron{n}

\newcommand{\ud}{\mathrm{d}}
\newcommand{\rv}{\mathbf{r}}
\def \br{\mathbf{r}}
\def \bx{\mathbf{x}}

\usepackage{mathtools}
\usepackage{tikz}
\makeatletter
\newcommand\mathcircled[1]{%
    \mathpalette\@mathcircled{#1}%
}
\newcommand\@mathcircled[2]{%
    \tikz[baseline=(math.base)] \node[draw,circle,inner sep=1pt] (math) {$\m@th#1#2$};%
}
\makeatother
\DeclareUnicodeCharacter{2264}{-}
\DeclareUnicodeCharacter{2265}{-}
\begin{document}

\preprint{AIP/123-QED}

\title[]{Comparing correlation components and approximations in Hartree-Fock and Kohn-Sham theories via an analytical test case study}
\author{Sara Giarrusso}
\author{Aurora Pribram-Jones}%
 \email{apj@ucmerced.edu}
\affiliation{ 
Department of Chemistry and Biochemistry, University of California Merced, 5200 North Lake Rd. Merced, CA 95343, USA
}%

\date{\today}

\begin{abstract}
The asymmetric Hubbard dimer is a model that allows for explicit expressions of the Hartree-Fock (HF) and Kohn-Sham (KS) states as analytical functions of the external potential, $\Delta v$, and of the interaction strength, $U$. We use this unique circumstance to establish a rigorous comparison between the individual contributions to the correlation energies stemming from the two theories in the $\{U, \,\Delta v\}$ parameter space. Within this analysis of the Hubbard dimer, we observe a change in the sign of the HF kinetic correlation energy, compare the indirect repulsion energies, and derive an expression for the `traditional' correlation energy, i.e. the one that corrects the HF estimate, in a pure site-occupation function theory spirit [Eq.~\eqref{eq:EchfHD}]. Next, we test the performances of the  Liu-Burke and the Seidl-Perdew-Levy functionals, which model the correlation energy based on its weak- and strong-interaction limit expansions and can be used for both the traditional and the KS correlation energies. Our results show that, in the Hubbard dimer setting, they typically work better for the HF reference, despite having been originally devised for KS. These conclusions are somewhat in line with prior assessments of these functionals on various chemical data sets. However, the Hubbard dimer model allows us to show the extent of the error that may occur in using the strong-interaction ingredient for the KS reference in place of the one for the HF reference, as has been carried out in most of the prior assessments.
\end{abstract}

\maketitle

\section{Introduction and theoretical background}\label{sec:intro}
Electronic structure problems in chemistry  can be addressed by means of  
an ever-increasing variety of methods, inviting both explicit and implied comparisons. 
The various methods are sometimes classified into the wavefunction-based category or into the density functional theory (DFT) one. 
A large part of this work is devoted to comparing the ``primal'' wavefunction method, i.e. Hartree-Fock (HF), and the most popular flavor of DFT, i.e. Kohn-Sham (KS).

These methods have been known for a very long time and, in recent years, the number of works where they are combined together has been increasing.~\cite{ShaTouSav-JCP-11, GhoVerCraGagTru-CR-18, VucSonKozSimBur-JCTC-19} However, systematic comparison of the two, contrasting their formal properties and guiding new approaches that combine them, is quite a hard task. A fair comparison, such as the one done in Ref.~\onlinecite{GriSchBae-JCP-97} for three simple diatomic molecules, requires the calculation of extremely accurate \textit{ab initio} wavefunctions from which the exact KS quantities (wavefunction, XC energy, etc) may be constructed. And this procedure is system-specific: 
it has to be repeated for any system for which one wishes to investigate how the two methods compare to one another, complicating systematic studies. Furthermore, the effect of the basis set used can hardly be identical in the HF and the KS states, introducing errors in the comparison. We bypass these disadvantages by adopting a radically simple model system: the asymmetric Hubbard dimer. In this model, both the HF and the KS states can be constructed analytically.

Before reviewing the model (Sec.~\ref{sec:HDintro}) as well as the two methods side by side (Sec.~\ref{sec:HF&KSintro}), let us introduce the usual non-relativistic Hamiltonian expression considered in electronic structure calculations,
 \begin{equation}\label{eq:ESham}
 \hat{H}=\hat{T}+\hat{V}_{ee} + \hat{V},
 \end{equation}
where $\hat{T}=-\sum_i^N\frac{\nabla_{i}^2}{2}$ is the kinetic energy operator, $ N $ is the number of particles in the system, $ \hat{V}_{ee} $ represents the Coulomb interaction between all electron pairs, and $ \hat{V} = \sum_i^N v (i)$ is the $N$-particle sum of the external potential, (typically) given by the positive field of the nuclei, felt by each electron.
Let $|\Psi\rangle $ be the wavefunction which solves the Schr\"odinger equation defined by $  \hat{H} $ with the lowest eigenvalue $E$.  
The solution, $|\Psi\rangle $, is notoriously hard to find, as it depends on the spatial and spin variables of each particle, i.e. $|\Psi\rangle =\Psi (\mathbf{x}_1, \cdots, \mathbf{x}_N)$, with $\mathbf{x}_i = \br_i \sigma_i$, the spatial and spin coordinates. On the other hand, the ground-state (GS) electron probability density, or just (electron) density,
\begin{equation}\label{eq:ron}
\ron (\br) :=  \langle \Psi |\hat{n} |\Psi\rangle, 
\end{equation}
where $\hat{n} = \sum_i^N \delta(\br_i - \br)$ and the Dirac brakets $ \langle \dots | \dots  \rangle $ stand for $ \int \ud\mathbf{x}_1 \cdots \ud \mathbf{x}_N  $, with $ \int \ud \bx = \sum_\sigma \int \ud \br $, is a much simpler mathematical object. 

Furthermore, if the ground state is unique, there is a bijective mapping between the wavefunction $\Psi$ and the external potential $v$, and by virtue of Eq.~\eqref{eq:ron}, also between $\ron$ and $v$, so that the expectation value of a suitable operator $\hat{A}$ evaluated on the GS wavefunction is also a functional of the GS density:
\begin{align}\label{eq:Aron}
A[\ron]=\langle \Psi[\ron]|\hat{A}|\Psi[\ron] \rangle. 
\end{align}

\subsection{The asymmetric Hubbard dimer}\label{sec:HDintro}
The general $N$-site Hubbard model was originally studied to describe the correlation effects in partially-filled narrow energy bands in solids.~\cite{Hub-PRSL-63, LieWu-PRL-68, Mon-book-92} It has gradually been used in the most diverse sceneries of physics and chemistry and is now often used as a playground to test new computational methods or concepts.~\cite{TheBucEicRugRub-JCTC-18, LacMar-PRL-20, MarBurLoo-JPCM-21}
Its two-site asymmetric version is relevant in the context of density functional theory,~\cite{CarFerSmiBur-JPCM-15, CohMor-PRA-16, YinBroLor-PRB-16} or Site Occupation Function Theory (SOFT) as is called in the lattice setting, and its offshoots (time-dependent DFT,~\cite{CarFerMaiBur-EPJB-18} density embedding theory,~\cite{SenTsuRobFro-MP-17} ensemble DFT~\cite{DeuMazFro-PRB-17} and thermal DFT~\cite{SmiPriBur-PRB-16}). Its simplicity allows a detailed, controlled and not rarely analytical exploration of the quantities of interest in these fields. 

The two-site Hubbard model Hamiltonian reads
\begin{equation}\label{eq:HDham}
\mathcal{\hat{H}}=\mathcal{\hat{T}}+\mathcal{\hat{U}}+\mathcal{\hat{V}}
\end{equation}
where
\begin{eqnarray}
\mathcal{\hat{T}}&=& -t \sum_\sigma \left(\hat{a}_{0\sigma}^\dagger\hat{a}_{1\sigma}+\hat{a}_{1\sigma}^\dagger\hat{a}_{0\sigma}\right)\\
\mathcal{\hat{U}}&=&U \sum_{i=0,1} \hat{n}_{i \uparrow}\hat{n}_{i \downarrow}  \label{eq:Uhat}\\
\mathcal{\hat{V}}&=&\sum_{i=0,1} v_i \hat{n}_i,
\end{eqnarray}
 $\hat{a}^\dagger,\, \hat{a} $ are the usual creation and annihilation operators, 
 $\sigma= \uparrow, \downarrow$ labels the spin of the particles, $i=0,1 $ labels the two sites, and $\hat{n}_{i\sigma} = \hat{a}_{i\sigma}^\dagger\hat{a}_{i\sigma}$ and $\hat{n}_i= \hat{n}_{i\sigma}+\hat{n}_{i\overline{\sigma}}$ (with $\overline{\sigma}$ being the spin opposite to $ \sigma $) are the occupation operators.
The parameters appearing in the Hamiltonian -- $t$, $U$, and $\{ v_i \}$ -- determine the aptitude of the particles to hop on the other site, the strength of the repulsion between particles, and their attraction to each site, respectively. In this sense, each term in the lattice Hamiltonian mimics the action of each term in the electronic Hamiltonian [Eq.~\eqref{eq:ESham}]. 

The eigenstates corresponding to Eq.~\eqref{eq:HDham} are fully determined by the reduced variables $u=\frac{U}{2\,t}$,  and $\delta v=\frac{\Delta v}{2\, t}$, with $\Delta v = v_1-v_0 $. Thus, we set $ t=1/2$ throughout the paper, as is customary.~\citep{CarFerSmiBur-JPCM-15}
Furthermore, we constrain the expectation value of the occupation operators on each site, $ n_i $, to add up to two (i.e., $  n_0+n_1= 2 $), and we consider only the states with $S_z =0$.
 Therefore, the Fock space reduces to three-dimensions and can be represented by the basis  $ | 0\uparrow 0\downarrow\rangle,|1\uparrow1\downarrow \rangle $ and the antisymmetric combination of singly occupied sites, $ \frac{1}{\sqrt{2}} \left( | 0\uparrow 1\downarrow\rangle-|0\downarrow 1\uparrow\rangle\right)$.  The associated Schr\"odinger equation can be solved analytically by finding the roots of a cubic polynomial, and all the quantities of interest can be compactly expressed by trigonometric formulas. 
  Note that, whereas we can generally express how the occupation difference $ \Delta n = n_1 -n_0 $ depends on $ U $ and $ \Delta v $, the inverse mapping (i.e., $ \Delta n \rightarrow \Delta v $) is not analytical.

We stress two fundamental features of $ \mathcal{\hat{H}} $, which set it apart from the electronic Hamiltonian of Eq.~\eqref{eq:ESham}: the first one is that the lack of a (second-order) derivative with respect to the particle space variable significantly alters the meaning of ``kinetic energy" in the quantum context (no Heisenberg principle, wave-particle duality and so on). In fact, the expectation value of the hopping operator is negative.
The second is that the two-body interaction in the lattice model is defined only between particles with opposite spin, a relevant difference from electrons, which interact with one another regardless of their spin. Therefore, the mean-field term [Eq.~\eqref{eq:EH} below] in the Hubbard model is free of the self-interaction error. Similarly, the exchange energy [Eq.~\eqref{eq:Ex} below], which specifically accounts for the interaction among particles of same spin, is exactly zero. Correspondingly, in this work, we shall compare the two theories, HF and KS, only in their correlation energy contributions.
To set the stage for this comparison we review the two theories in general terms in the next section.

\subsection{Hartree-Fock and Kohn-Sham methods} \label{sec:HF&KSintro}
According to the Hartree-Fock method, the expectation value of the Hamiltonian in Eq.~\eqref{eq:ESham} is minimized in the space of Slater determinants, $ \Phi  := \sum_{P}(-1)^P\psi_{P(1)}(\mathbf{x}_1)\cdots\psi_{P(N)}(\mathbf{x}_N)$, where the $\psi_n(\mathbf{x})$ are single-particle wavefunctions and the spatial and spin coordinates are considered separable, i.e. $\psi_n(\mathbf{x}) \equiv \phi_n(\rv) s_{n}(\sigma)$, while the index $ P $ lists all possible permutations. 
Its ground state is then given by
\begin{equation}\label{eq:HFwf}
| \Phi^\hf \rangle =\text{argmin}_{\Phi} \langle \Phi| \hat{H}|\Phi \rangle.
\end{equation}
The corresponding ground-state density, $ \ron^\hf (\rv)= \sum_\sigma\sum_i^N |\psi_i^\hf(\mathbf{x})|^2$, typically differs from the interacting one [Eq.~\eqref{eq:ron}]. 
The HF approximation to the GS energy is 
\begin{equation}\label{eq:HFappE}
E^\hf := \langle \Phi^\hf | \hat{H} |\Phi^\hf  \rangle
\end{equation}
and, by virtue of the variational principle, $  E^\hf  \geq E $.
Their difference is usually referred to as simply the ``correlation energy" $ E_c $; we shall however label it  the Hartree-Fock correlation energy, $  E_c^\hf$, to distinguish it from the KS one.
It is defined as the difference between the GS energy and its HF approximation, $ E^\hf $, and
 consists of the following individual contributions
\begin{equation}\label{eq:HFEc}
E_c^\hf= T_c^\hf + U_c^\hf + V_c^\hf,
\end{equation}
where
\begin{eqnarray}
T_c^\hf & := &  T[\ron] - T^\text{SD}[\{\psi^\text{HF}_i \}]  \label{eq:Tc} \\
U_c^\hf & := & V_{ee}[\ron] -  U_H[\ron^\hf] - E_x[\{\psi^\text{HF}_i \}]   \label{eq:Uc} \\
V_c^\hf & := & V[\ron] -V[\ron^\hf] \label{eq:Vc}. 
\end{eqnarray}
Here, the first terms on the right hand side  of Eqs.~\eqref{eq:Tc}-\eqref{eq:Vc} are applications of Eq.~\eqref{eq:Aron}, while 
$ T^\text{SD}[\{\psi_i \}] = \frac{1}{2} \sum_i^N \int |\nabla_\br \psi_i(\mathbf{x})|^2 \ud \bx$ is the kinetic energy as evaluated on a Slater determinant, 
\begin{equation}\label{eq:EH}
    U_H[\ron]:=\frac{1}{2}\int \int \frac{\ron(\br) \ron(\br')}{|\br-\br'|} \ud \br \ud \br'
\end{equation}
is the mean field repulsion energy,  and 
\begin{equation}\label{eq:Ex}
  E_x[\{\psi_i \}] = -\frac{1}{2} \sum_{i, j}^{N} \int  \int  \frac{\psi_i^* (\bx) \psi_j^*(\bx) \psi_i (\bx') \psi_j (\bx')}{|\br -\br'|} \ud \bx \ud \bx'  
\end{equation}
is the exchange energy, which comes from evaluating the interaction operator on a Slater determinant and subtracting the mean-field term. Moreover, the external potential energy functional is a simple explicit functional of the density,
$V[\ron]:= \int v(\br) \ron (\br) \ud \br$.

In the Kohn-Sham formulation of DFT, the full Hamiltonian is set aside and only
the kinetic energy operator is minimized over all antisymmetric $N$-particle wavefunctions. However,  the minimization is performed under the constraint of a fixed density. The resulting density functional is known as the Kohn-Sham kinetic energy functional, $T_s[\ron]$
\begin{equation}\label{eq:Tsron}
T_s[\ron] := \min_{\Psi \to \ron} \langle \Psi | \hat{T} | \Psi \rangle.
\end{equation}
The minimizing wavefunction is expected to be a Slater Determinant, as there are no two-body operators entering the minimization, although there are cases in which
the single Slater determinant description cannot deliver the prescribed density. \cite{SchGriBae-TCA-98, Lee-AQC-03, GieBae-JCP-10} If we neglect such cases, then $ T_s[\ron] \equiv T^\sd[\{\psi_i^\ks \}]$, where the KS orbitals $ \{\psi_i^\ks \}  $ are the one-particle functions cast in the KS Slater determinant, $ \Phi^\ks $, which is the minimizer of the search on the right hand side of Eq.~\eqref{eq:Tsron}. The KS orbitals are clearly functionals of the density, though in an implicit and highly non-trivial way. Conversely, the interacting density is easily written in terms of the KS orbitals as $ \ron(\rv)= \sum_\sigma\sum_i^N |\psi_i^\ks(\mathbf{x})|^2$.
The correlation energy according to KS-DFT, $ E_c^\ks $, is given by
\begin{equation}\label{eq:KSEc}
E_c^\ks = T_c^\ks + U_c^\ks,
\end{equation}
where $  T_c^\ks $ and $ U_c^\ks $ look formally identical to Eqs.~\eqref{eq:Tc} and \eqref{eq:Uc} respectively, with the ``non-interacting" pieces having the KS orbitals $ \{\psi_i^\ks \} $ and the interacting density as input, rather than the HF quantities.
The missing external potential contribution in Eq.~\eqref{eq:KSEc} compared to Eq.~\eqref{eq:HFEc} is a result of the KS density being, by construction, equal to the interacting one. This is at the root of KS-DFT being an exact treatment, rather than an approximation strategy like Hartree-Fock.
The matching between the density of the ``non-interacting"  auxiliary system and that of the interacting target system is enforced by means of an effective external potential, called the KS potential, $ v_s $. 
To see the relation between said potential and Eq.~\eqref{eq:KSEc}, one may decompose it into 
\begin{equation}
v_s = v + v_H + v_{xc},
\end{equation}
where $v$ is the external potential of the target problem [Eq.~\eqref{eq:ESham}], $ v_H $, is the Hartree potential defined as the functional derivative of $ U_H[\ron] $, i.e. $v_H [\ron] (\br)=\int \frac{ \ron(\br')}{|\br-\br'|} \ud \br'$, while $v_{xc}$ is the so-called exchange-correlation (XC) potential. This corresponds to the functional derivative of the XC energy
\begin{equation}\label{eq:vxcdef}
v_{xc}[\ron_0]= \frac{\delta\, E_{xc}^\ks [\ron]}{\delta \ron}  \Big|_{\ron=\ron_0},
\end{equation}
with $E_{xc}^\ks [\ron] =E_c^\ks[\ron] + E_x[\{\psi_i^\ks \} [\ron]] $.\\
As $E_{xc}^\ks$ is not known in general, this term has to be approximated in actual KS-DFT calculations. Though we have access to the numerically exact quantity for our Hubbard dimer, we present one possible route to build approximations for it in the next section.

\subsection{Approximations from the adiabatic connection framework}
A quite powerful and long-established tool to construct approximation for the XC energy in KS-DFT is represented by 
the density-fixed adiabatic connection formalism.~\cite{HarJon-JPF-74, GunLun-PRB-76, LanPer-SSC-75, Lan-PRL-84} 
According to this formalism, a parameter $\l$ is used to tune the strength of the interaction operator in the Hamiltonian~\eqref{eq:ESham} while keeping the density fixed,
under the assumption that the density is $v$-representable for all $\lambda$, i.e.:
 \begin{equation}\label{eq:Hldft}
\hat{H}_{\lambda}^\ks= \hat{T} + \lambda \hat{V}_{ee} + \hat{V}^{\lambda},
\end{equation}
where $\hat{V}^{\lambda}=\sum_i^N v^{\lambda}(\br_i)$ and $v^\lambda$ is the Lagrange multiplier that keeps the density fixed at each $\lambda$.
One can then show that
 \begin{equation}
 E_{xc}^\ks [\ron] = \int_0^1 W_\l^\ks [\ron] \ud \l,
 \end{equation}
with the AC integrand defined as
\begin{equation}\label{eq:Wldft}
W_\l^\ks [\ron] := \langle \Psi_{\lambda}^\ks[\ron] | \hat{V}_{ee}|\Psi_{\lambda}^\ks[\ron]\rangle-U_H[\ron]
\end{equation}
and with $\Psi_\l^\ks$ the ground state of the $\l$-dependent Hamiltonian~\eqref{eq:Hldft} at each $\l$.\\
The exact behaviour of $W_\l^\ks$ is known locally in the two limits $\l \to 0$~\cite{GorLev-PRB-93, GorLev-PRA-94} and $\l \to \infty$\cite{Sei-PRA-99, GorVigSei-JCTC-09}:
\begin{eqnarray}
W_{\l\to 0}^\ks [\ron] & = &E_x [\{\psi_i^\ks [\ron] \}]+ \sum_{\text{n}=2}^\infty \text{n}\,E_c^{\text{GLn}} \l^{\text{n}-1}\label{eq:lto0} \\
W_{\lambda\rightarrow\infty}^\ks[\ron] & = & W_\infty^\ks [\ron] + O\left( \l^{-\frac{1}{2}} \right)\label{eq:ltoinfty},
\end{eqnarray}
with $E_c^{\text{GLn}}$ the $n$th-order Gorling-Levy (GL) correlation energy coefficients,~\cite{GorLev-PRB-93, GorLev-PRA-94} and $W_\infty^\ks$ the minimal repulsion energy in a given density removed of its mean-field part.\cite{Sei-PRA-99, SeiGorSav-PRA-07}

Models for the KS-DFT XC energy based on interpolating between the weak- and the strong-interaction expansions~\eqref{eq:lto0} and~\eqref{eq:ltoinfty} are called
Adiabatic Connection Interaction Interpolations (ACIIs) or Adiabatic Connection Methods (ACMs). As a matter of fact, these approximations, developed within
KS-DFT, have been successfully used
with HF ingredients as a correction
to the HF energy.~\cite{FabGorSeiDel-JCTC-16, VucGorDelFab-JPCL-18, GiaGorDelFab-JCP-18, DaaFabSalGorVuc-JPCL-21} Such practice began from the simple heuristic observation that using ACMs on HF ingredients gave consistently better results than using them on KS ones.~\cite{FabGorSeiDel-JCTC-16}

While adaptations of the adiabatic connection approach to wavefunction methods had already begun to appear (see Ref.~\onlinecite{Per-IJQC-18} and references therein), the key factor needed to justify the use of ACMs with different reference states was the boundedness of the leading coefficient in the strong-interaction expansion of the corresponding AC integrand. 
In the case of the AC with the HF state as reference, such boundedness was shown only more recently.~\cite{SeiGiaVucFabGor-JCP-18} 

Within this other adiabatic connection framework, the $\l$-dependent Hamiltonian reads
\begin{equation}\label{eq:MPac}
\hat{H}_\l^\hf = \hat{T} +\hat{V}_\hf +\hat{V} +\l \left( \hat{V}_{ee}-\hat{V}_\hf\right),
\end{equation}
where $ \hat{V}_\hf =\sum_{i,j}^N \left(\hat{J}_j ^\hf(\bx_i) - \hat{K}_j^\hf(\bx_i) \right)  $, 
\begin{equation}\label{eq:Ji}
\hat{J}_i^\hf (\bx) = \int \frac{|\psi_i^\hf (\bx)|^2}{|\br -\br'|} \,\ud \bx'
\end{equation}
and
\begin{equation}\label{eq:Ki}
\hat{K}_i^\hf (\bx) \phi (\bx) = \psi_i^\hf(\bx) \int \frac{\psi_i^{\hf \ast}(\bx') \, \phi (\bx')}{|\br -\br'|}\, \ud \bx'.
\end{equation}
Similarly to the DFT case, one can show that
\begin{equation}
E_c^\hf = \int_0^1 W_{\l}^\hf \ud \l,
\end{equation}
with the adiabatic connection integrand, $ W_{\l}^\hf $, defined as
{\small
\begin{equation}\label{eq:Wlmp}
W_{\l}^\hf := \langle \Psi_\l^\hf |\hat{V}_{ee} - \hat{V}_\hf |\Psi_\l^\hf  \rangle +  c_0^\hf [\ron^\hf],
\end{equation}}
and $ c_0^\hf [\ron^\hf]=U_H[\ron^\hf]+E_x[\{ \psi_i^\hf\}]  $. Note that we adopt a slightly different notation than the one in Ref.~\onlinecite{SeiGiaVucFabGor-JCP-18}  for the adiabatic connection integrand: in definition~\eqref{eq:Wlmp}, $W_{\l}^\hf$ is gauged to go to zero when $\l=0$ rather than to $E_x [\{\psi_i^\hf \}]$ as in the original paper.

The small and large $ \l $ expansions of $W_{\l}^\hf$ give 
\begin{eqnarray}\label{eq:WsmallHF}
W_{\l\to 0}^\hf & = &\sum_{\text{n}=2}^\infty \text{n}\,E_c^{\text{MPn}} \l^{\text{n}-1},\\
W_{\l\to \infty}^\hf & = & W_\infty^\hf + O\left( \l^{-\frac{1}{2}} \right)\label{eq:WlargeHF},
\end{eqnarray}
with $ E_c^{\text{MPn}}$ the $n$th-order M\o ller-Plesset (MP) correlation coefficients and 
\begin{equation}
W_\infty^\hf = E_{el}[\ron^\hf] + E_x[\{\psi_i^\hf\}],
\end{equation}
with
{\small
\begin{equation}\label{eq:EelHF}
	E_{\rm el}[\ron]\equiv \min_{\{\rv_1\dots\rv_N\}}\left\{\sum_{i,j>i}^{N}\frac{1}{|\rv_i-\rv_j|}-\sum_{i=1}^N v_{\rm H}(\rv_i;[\ron])+U_H[\ron]\right\}
\end{equation}}
the minimum total electrostatic energy of $N$ equal classical point charges $(-e)$ in a positive background with continuous charge density $(+e)\ron(\rv)$.\cite{SeiGiaVucFabGor-JCP-18}

The ACMs strategy of interpolating between the weak- and the strong-interaction expansions  of the desired AC integrand has the major merit of providing an all-order resummation of the perturbation series coefficients by encompassing also the strong-interaction information. This avoids difficulties such as slowly convergent or divergent series (see, e.g., a discussion of the shortcomings associated with MP theory in quantum chemistry in Ref.~[\onlinecite{purplebible}]).

In this work, we test two ACMs that depend on three ingredients: $E_x$, $E_c^\text{PT2}$, where `PT' stands for `Perturbation Theory', and $W_\infty$. Note that, while the ingredient $E_x$ is formally exactly the same regardless of the reference used (HF or KS) and only the input quantities change (i.e. HF or KS orbitals), the ingredients $E_c^\text{PT2}$ and $W_\infty$ correspond respectively to $E_c^\text{MP2}$ and $W_\infty^\hf$ for the HF reference and $E_c^\text{GL2}$ and $W_\infty^\ks$ for the KS reference.\\
Specifically, the functionals considered in this work are the Liu-Burke (LB),\cite{LiuBur-PRA-09}
{\small{\begin{equation}\label{eq:LB}
  E_c^\text{LB}  =-\frac{\tilde{W}_\infty^2 \left(\sqrt{\frac{20 W_0'}{\tilde{W}_\infty}+25}-5\right)}{4 W_0'}-\frac{5
   \tilde{W}_\infty^2}{8 W_0'+10 \tilde{W}_\infty}+\tilde{W}_\infty,
\end{equation}}}
and the Seidl-Perdew-Levy (SPL),~\cite{SeiPerLev-PRA-99} 
{\small{\begin{equation}\label{eq:SPL}
  E_c^\text{SPL} =\tilde{W}_\infty-\frac{\tilde{W}_\infty^2 \left(\sqrt{1+\frac{2 W_0'}{\tilde{W}_\infty}}-1\right)}{W_0'}, 
\end{equation}}}%
where in both equations we have used $W_0'=2\,E_c^\text{PT2}$ and $\tilde{W}_\infty=W_\infty-E_x$.

\section{Comparison between $E_c^\hf$ and $E_c^\ks$}
Both the HF and the KS states that correspond to the interacting problem introduced in Eq.~\eqref{eq:HDham} can be constructed analytically.~\citep{CarFerSmiBur-JPCM-15} 
This quite rare (if not unique) circumstance allows us to establish a detailed comparison between the two theories (see Figs~\ref{fig:Tcksvshf}, \ref{fig:Ucksvshf}, \ref{fig:Vchf} and \ref{fig:Ecksvshf}).
Moreover, we discuss both these methods from a site-occupation function theory standpoint (see Figs~\ref{fig:inversionproblem}, \ref{fig:NNinset} and~\ref{fig:Ecvsn} and Eq.~\eqref{eq:EchfHD}).  This is quite usual for KS theory but rather uncommon for HF. Before showing the results for the HF and KS correlation energies and their individual contributions (Sec.~\ref{sec:indcons}), we review such methods as applied to
the Hubbard dimer.

\subsection{Mean-field solutions: an overview}
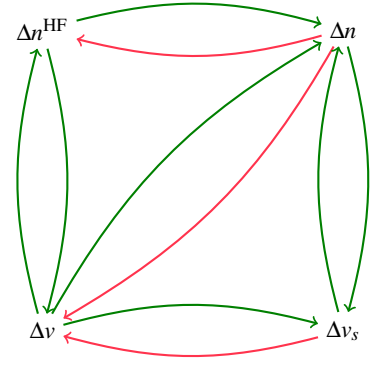
\begin{figure}
\begin{tikzpicture}
\node (nhf) at (0,4) {$\Delta n^\hf$};
\node (n) at (4,4) {$\Delta n$};
\node (v) at (0,0) {$\Delta v$};
\node (vs) at (4,0) {$\Delta v_s$};
\path[->,thick, go] (nhf) edge [bend left=15] (n);
\path[->,thick, nogo] (n) edge [bend left=15] (nhf);
\path[->,thick, go] (v) edge [bend left=15] (nhf);
\path[->,thick, go] (nhf) edge [bend left=15] (v);
\path[->,thick, go] (v) edge [bend left=15] (n);
\path[->,thick, nogo] (n) edge [bend left=15] (v);
\path[->,thick, go] (v) edge [bend left=15] (vs);
\path[->,thick, nogo] (vs) edge [bend left=15] (v);
\path[->,thick, go] (vs) edge [bend left=15] (n);
\path[->,thick, go] (n) edge [bend left=15] (vs);%
\end{tikzpicture}
\caption{A graphical representation of the possible occupation and external potential functions.}
\label{fig:inversionproblem}
\end{figure}

In the non-interacting case, when $ U \equiv 0 $ in Eq.~\eqref{eq:HDham},  the solution to the Schr\"odinger equation is particularly simple. The GS occupation, which can be constructed from the non-interacting solution, is analytically invertible in terms of the potential and reads
\begin{equation}\label{eq:TBpot}
\Delta v = - \frac{\Delta n}{\sqrt{4 -\Delta n^2}}.
\end{equation}
In the Hubbard model community, this case is referred to as the ``tight-binding" problem.
As this case is relevant for the application of the KS method, and in order to distinguish the external potential (difference) pertaining to the target interacting problem from its non-interacting effective mapping, we relabel the external potential found from Eq.~\eqref{eq:TBpot} as ``$\Delta v_s$" (see also right edge of Fig.~\ref{fig:inversionproblem}).

Consider now the HF Hamiltonian for the Hubbard dimer: 
\begin{equation}\label{eq:HFham}
\mathcal{\hat{H}}^\hf = \mathcal{\hat{T}}+\sum_{i=0,1} \sum_\sigma U n_{i\overline{\sigma}}^\hf \hat{n}_{i\sigma} + \sum_{i=0,1} v_i \hat{n}_i.
\end{equation}
In contrast to Eq.~\eqref{eq:HDham}, in Eq.~\eqref{eq:HFham}, there is no interaction term and the repulsion is taken into account in a mean field fashion by the central term, $  U  n_{i\overline{\sigma}}^\hf$. On each site, the occupation with spin $ \sigma $ feels the repulsion generated by the spin-$ \overline{\sigma} $ occupation of the same site. We reiterate that, contrary to the usual continuum setting, here there is no repulsion between particles of the same spin. In Eq.~\eqref{eq:HFham}, this central term is reported as converged to the stationary point at which the mean field is generated by the HF occupation.

Furthermore, if we require that the spin up and spin down occupations are equal, we can substitute $n_{i\overline{\sigma}}^\hf=n_{i \sigma}^\hf=\frac{n_{i}^\hf}{2}$ in Eq.~\eqref{eq:HFham}, obtaining the corresponding restricted Hartree-Fock (RHF) Hamiltonian:
\begin{equation}\label{eq:HFhamTB}
\mathcal{\hat{H}}^\text{RHF} = \mathcal{\hat{T}}+\sum_{i=0,1}  \tilde{v}_i \hat{n}_i,
\end{equation}
with
\begin{equation}\label{eq:vtilde}
\tilde{v}_i = v_i + \frac{U}{2}n_i^\hf .
\end{equation}
In the following, we will always consider the RHF occupation resulting from Eqs.~\eqref{eq:HFhamTB}-\eqref{eq:vtilde}, referring to it as simply ``HF."

Note that, while the eigenvectors of Eq.~\eqref{eq:HDham} do not depend on the individual values, $v_i$, of the external potential but only on their difference $ \Delta v $, the energy does depend on the individual $v_i$.
One can also express the energy as a function of $\Delta v$ and a constant $c$ which represents the gauge choice, $c= v_0 + v_1$, and is typically set to zero.
However, setting $ v_0 +v_1=0 $ forces the $\tilde{v}_i$ of Eq.~\eqref{eq:vtilde} to give $\tilde{v}_0+\tilde{v}_1 = U$ (since $n_0^\hf+n_1^\hf = 2 $).
From Eq.~\eqref{eq:HFhamTB}, it follows that we can reconstruct the external potential as a function of the HF occupation by altering Eq.~\eqref{eq:TBpot} to include a mean field repulsion term and depend on $\Delta n^\hf$:
\begin{equation}\label{eq:deltavdeltan}
\Delta v = -\frac{U}{2} \Delta n^\hf -\frac{\Delta n^\hf}{\sqrt{4-\left( \Delta n^\hf\right)^2}}.
\end{equation}
The above function is also invertible and $\Delta n^\hf $  can be  expressed as a function of $ \{U,\Delta v\} $. We avoid reporting it here, but it can be found together with the other formulas and plots given in this work in a supporting notebook available for download.

In Fig.~\ref{fig:inversionproblem}, we visually summarize all the analytical pathways that connect the target interacting state to the two different ``non-interacting" reference states. For each reference state, we can construct the external potential from its corresponding  GS occupation and vice versa (as sketched in the left and right sides of the picture), whereas the connection between the two reference states work only in the direction from left to right and not the other way around. Indeed, if we could either construct the external potential of the interacting system from knowledge of the KS one (bottom arrow of the picture) or construct the KS occupation from knowledge of the HF one (upper arrow of the picture) we would find the external potential of the interacting system as a function of its GS occupation. 
As already mentioned, this problem has no analytical solution even in this extremely gaunt model system, except in the symmetric case, i.e. $\Delta v =0$, or in the limit where the interaction energy dominates over the kinetic/hopping one, i.e. $U \to \infty$.
\begin{figure}
 \begin{tabular}[c]{c}
 $\quad\quad\,${\begin{subfigure}{}
      \includegraphics[scale=0.26]{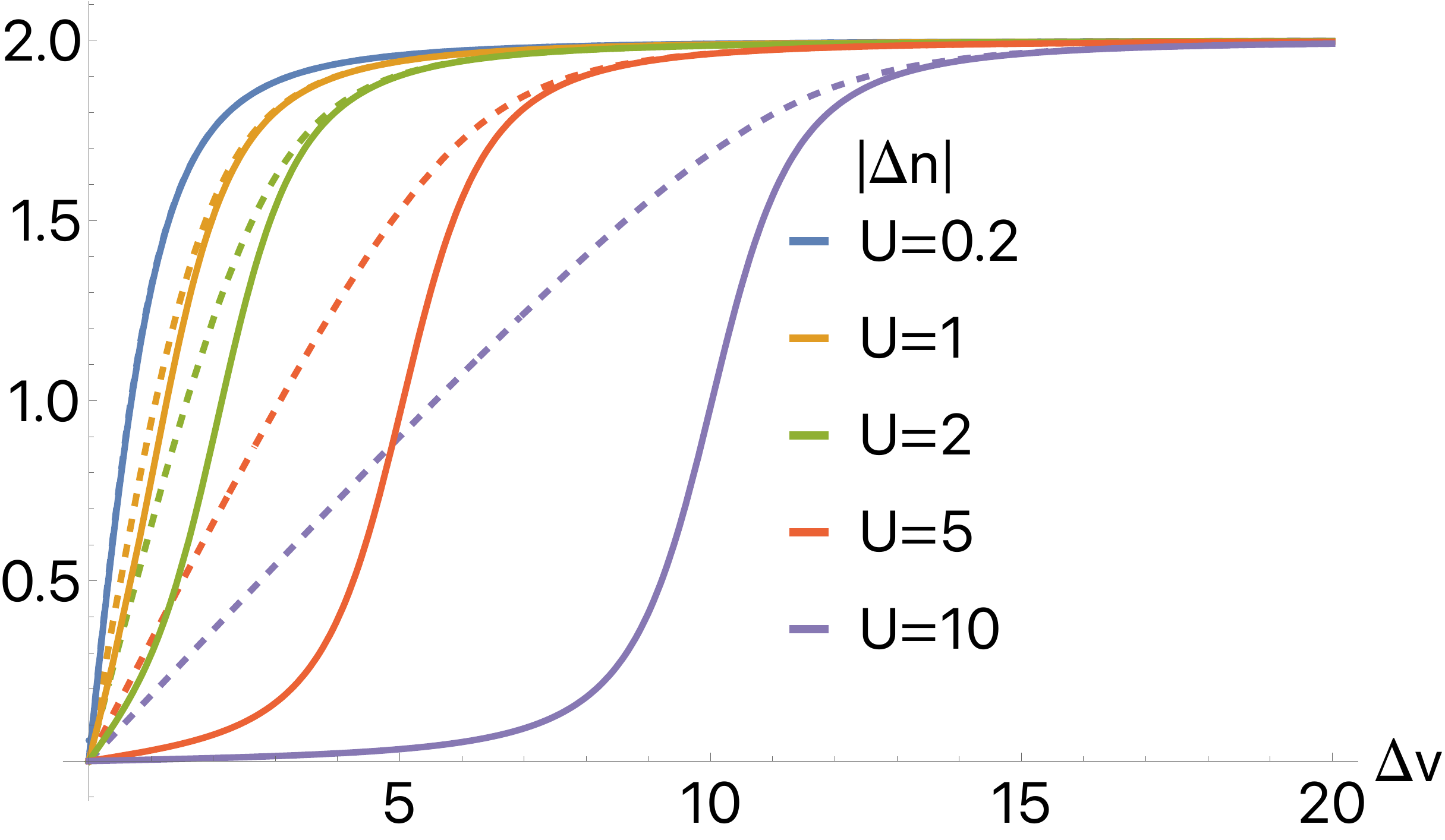}
    \end{subfigure}
} \\
 {\begin{subfigure}{}
 \includegraphics[scale=0.18]{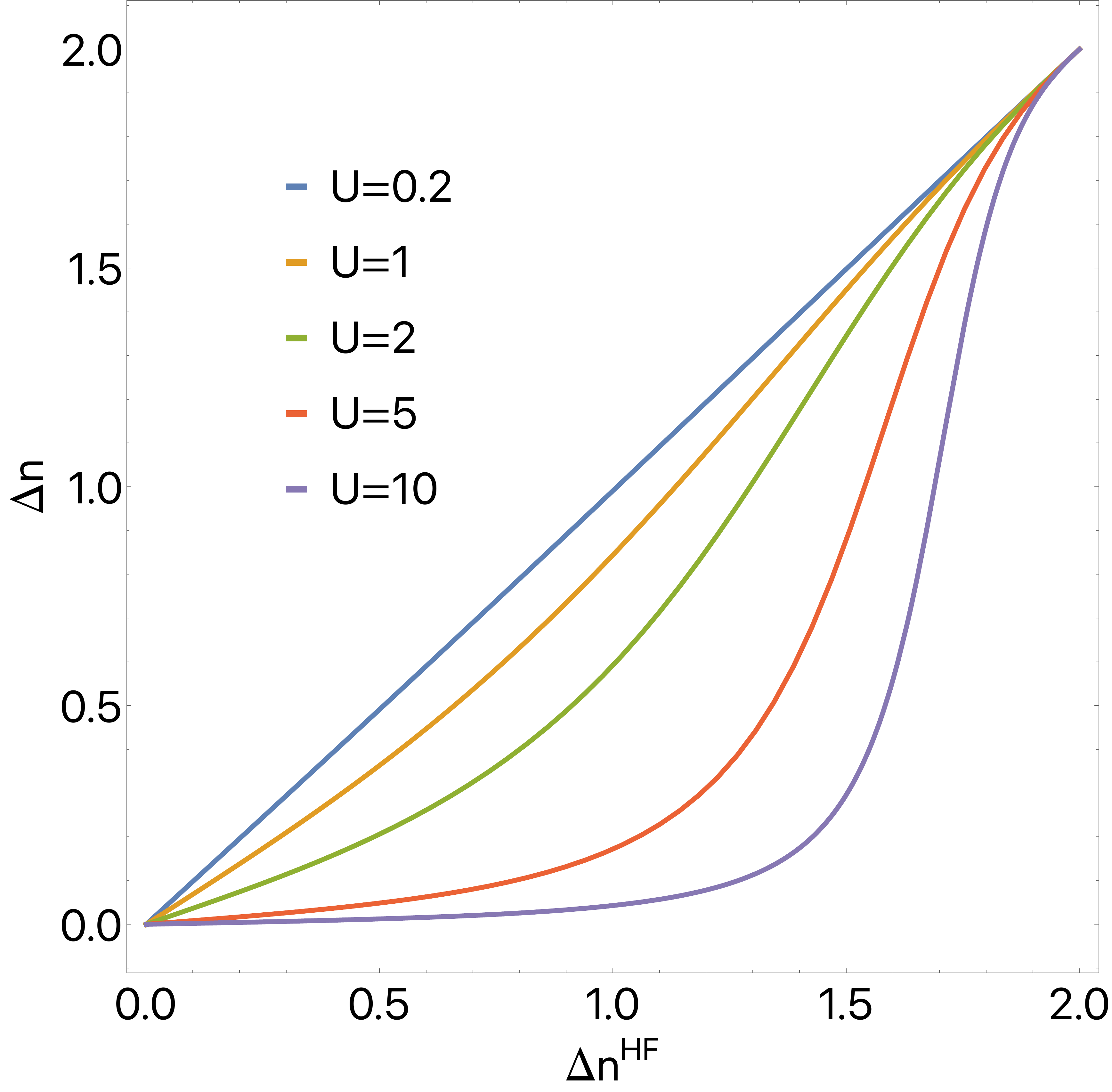}
 \end{subfigure}}
 \end{tabular}  
 \caption{Upper panel: site occupation differences for the interacting (solid) and the HF (dashed) systems as a function of the external potential difference, $ \Delta v $, for $ U=0.2,\,1,\,2,\,5,\,10 $. Lower panel: interacting site occupation difference, $ \Delta n $, as a function of the HF one, $ \Delta n^\hf $, for $ U=0.2,\,1,\,2,\,5,\,10 $.} 
\label{fig:NNinset}
\end{figure}

To conclude this section, in Fig.~\ref{fig:NNinset}, we plot the functions $ \Delta n$ and  $\Delta n^\hf $ along the external potential difference in the upper panel and against each other in the bottom panel.
  When $ U $ is small, e.g. $ U=0.2 $, the two occupations differ very slightly, while their difference increases for larger $ U $, as expected. The function $ \Delta n $ vs. $ \Delta v $, shown in the upper panel, can be quite flat for extremely large portions of its domain, on the other hand the function $ \Delta n $ vs. $ \Delta n^\hf $, on the bottom panel, appears to be much gentler at least for intermediate values of $ U $ (becoming non-analytical for $ U \to \infty $). Thus, from a numerical point of view, it is in general much more convenient to invert this latter relation and then use the function $ \Delta v $ vs. $ \Delta n^\hf $, rather than directly inverting the function $\Delta v $ vs. $ \Delta n $ (diagonal of Fig.~\ref{fig:inversionproblem}). We see that the two site occupations become equal in the symmetric limit, $ \Delta v =0 $, and asymptotically (i.e., for $ \Delta v \to \pm \infty$), where in both cases, the interplay between $ U $ and $ \Delta v $ vanishes.

\subsection{Individual contributions to the correlation energies}\label{sec:indcons}
\begin{figure}
\includegraphics[scale=0.28]{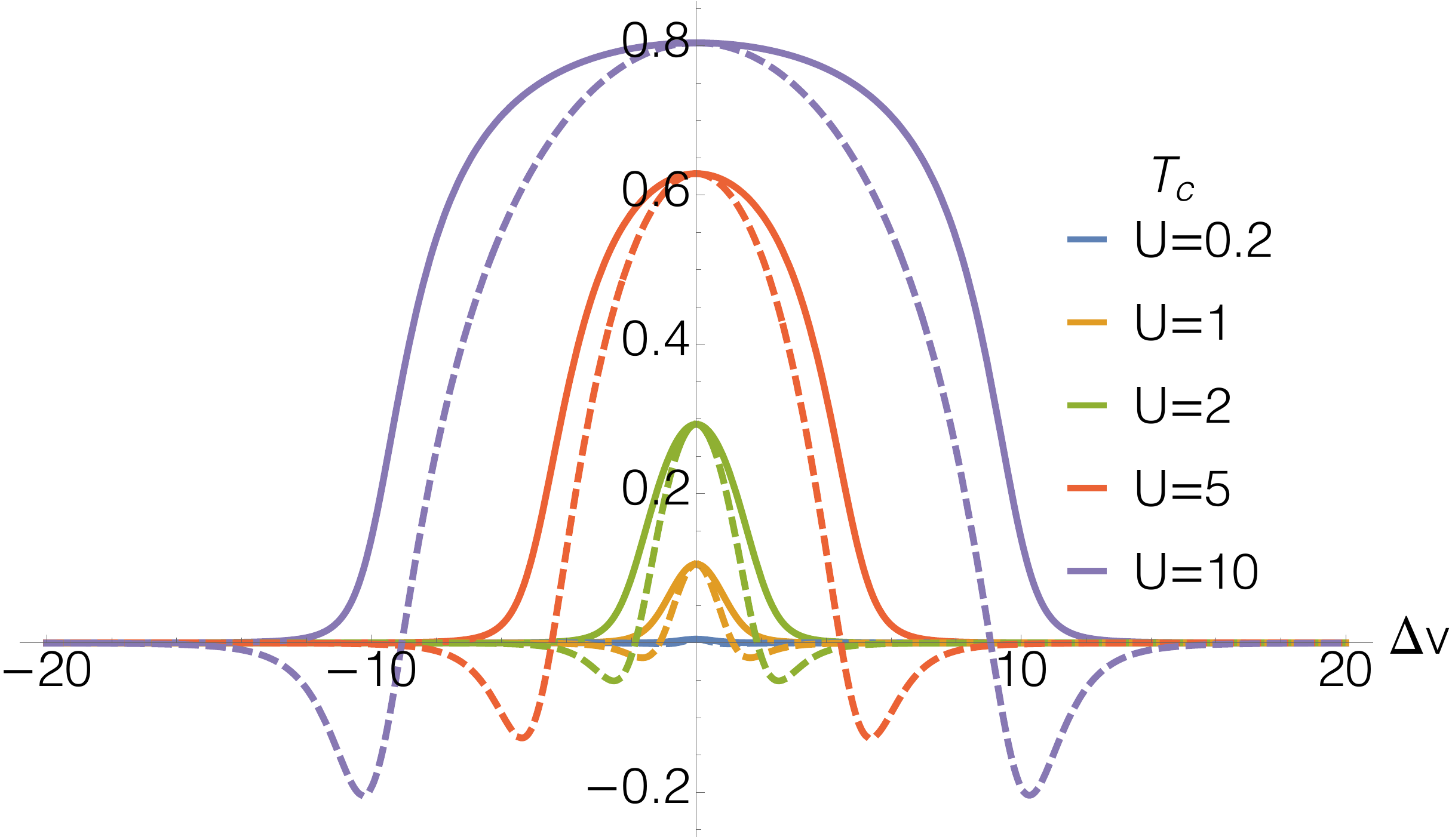}
\caption{Correlation kinetic energy contribution $T_c^\text{SD}$ (Eq.~\eqref{eq:Tc}) for the Hubbard dimer as a function of $ \Delta v $ and for $ U =0.2, 1, 2, 5, 10$. The dashed curve corresponds to $\text{SD}=\hf $ while the solid one to $\ks $.} 
\label{fig:Tcksvshf}
\end{figure}

In a spirit similar to that of Reference~\onlinecite{GriSchBae-JCP-97}, in this section we compare the individual contributions to the HF and the KS  correlation energies with one another. We begin by comparing kinetic correlation energies when we fix the external potential. By definition [Eq.~\eqref{eq:Tsron}], $ T_s $ is the minimal kinetic energy for a given density. However, as mentioned previously, when we solve a quantum problem for a given external potential, $ n^\hf \neq n $. It then becomes interesting to compare the correlation kinetic energy contribution in the two theories, $ T_c ^\hf$ and $T_c^\ks$ [Eq.~\eqref{eq:Tc}] as a function of the external potential difference,~$\Delta v $. This is done in Fig.~\ref{fig:Tcksvshf} 
for different $ U $ values. The dashed curve corresponds to $\text{SD}=\hf $ while the solid one to $\ks $ (as is the case in all the following figures in this section).  Note that $ T^\text{SD} [\Delta n]= -\sqrt{1-\left(\frac{\Delta n}{2} \right)^2} $ for any non-interacting reference state, so that actually $ T^\ks \equiv T^\hf $ as a function of a given site occupation. If however we consider the HF or KS kinetic energy for a particular external potential, we see that the inequality $ T^\ks \leq T^\hf $ still holds, becoming an equality when the interacting and the HF site occupations become equal (i.e. $ \Delta v =0 $ and $ \Delta v \to \infty $). Furthermore, we observe that, for each $ U $, there is a turning point in $|\Delta v|$ at which $ T_c^\hf \equiv 0$ and past which the HF hopping energy is higher than the interacting one, showing first evidence, to our knowledge, of $ T_c^\hf < 0$. 

 \begin{figure}
\includegraphics[scale=0.28]{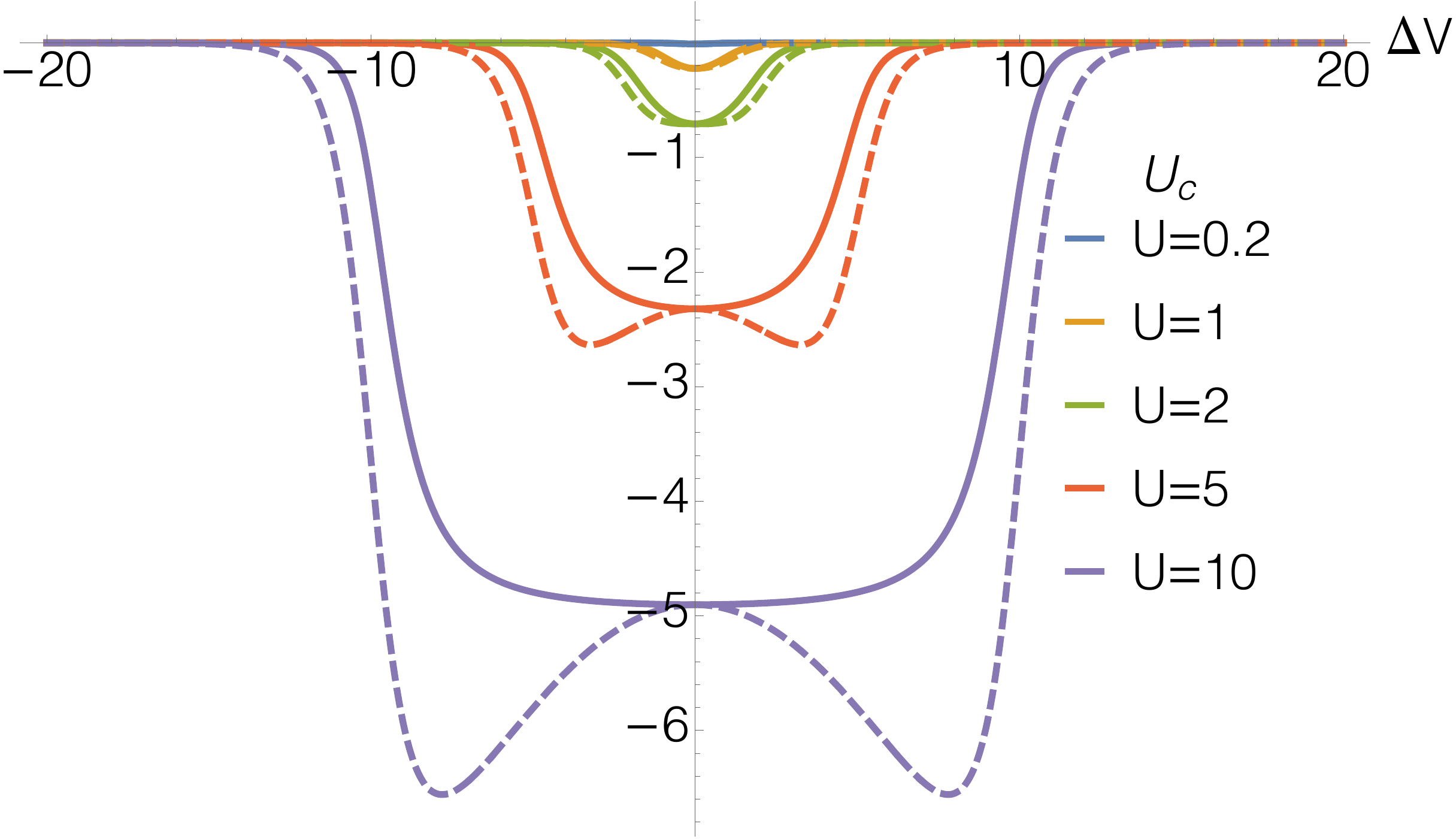}
\caption{Correlation Coulomb energy contribution $U_c^\text{SD}$ (Eq.~\eqref{eq:Uc}) for the Hubbard dimer as a function of $ \Delta v $ and for $ U =0.2, 1, 2, 5, 10$. The dashed curve corresponds to $\text{SD}=\hf $ while the solid one to $\ks $.} 
\label{fig:Ucksvshf}
\end{figure}
Moving on to comparison of $ U_c^\hf $ and $ U_c^\ks $ [Eq.~\eqref{eq:Uc}] in Fig.~\ref{fig:Ucksvshf}, we see that the discrepancy of this energy contribution is systematically larger (in magnitude) in the HF reference state than in the KS one. Note that $U_c^\hf$ contains also the correction coming from the Hartree term,
 $ U_{H,c}^\hf : =\frac{U}{2} \left( \left(\frac{\Delta n}{2} \right)^2 - \left(\frac{\Delta n^\hf}{2} \right)^2 \right) $. Quite interestingly, we find that the indirect Coulomb correlation energy defined as $ U_{ind,c}^\text{SD} := U_c^\text{SD} -  U_{H,c}^\text{SD}$ is exactly the same in the two reference states, meaning that $ U_{ind,c}^\hf \equiv U_{c}^\ks $. 
Therefore, the discrepancy observable in Fig.~\ref{fig:Ucksvshf} among the two methods is entirely due to the term $ U_{H,c}^\hf $, the mean field correction for the HF site occupation being different than the interacting one. Because the comparison is made with fixed external potential and the HF and KS site-occupations typically differ (see Fig.~\ref{fig:NNinset}), this means in turn that the dependence of $ U_{ind,c}^\text{SD}$ on the site-occupation in the two treatments is different.

\begin{figure}
\includegraphics[scale=0.28]{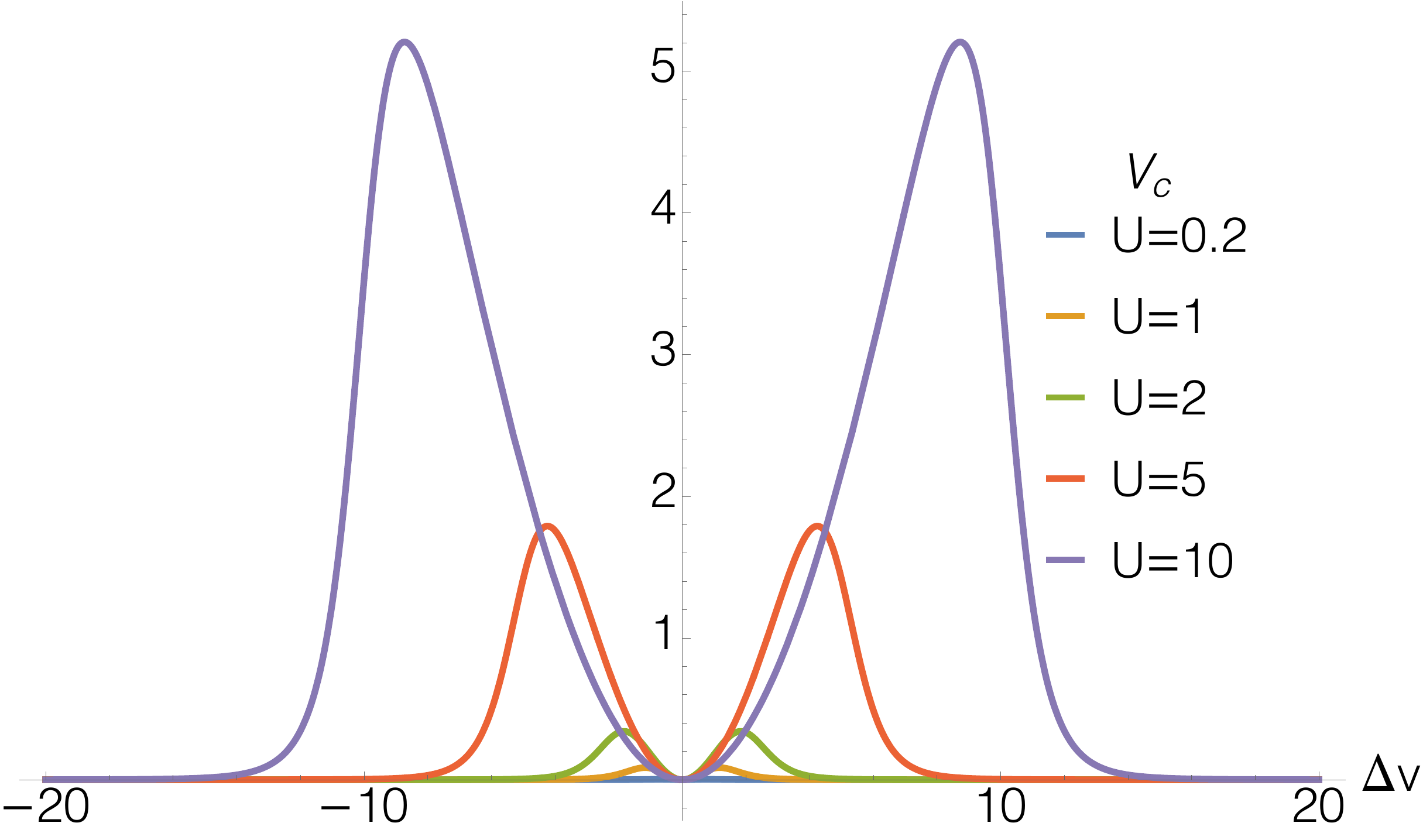}
\caption{Correlation potential energy contribution $V_c^\hf$ (Eq.~\eqref{eq:Vc}) for the Hubbard dimer as a function of $ \Delta v $ and for $ U =0.2, 1, 2, 5, 10$. This contribution is exactly zero for the $\ks $ reference.} 
\label{fig:Vchf}
\end{figure}

\begin{figure}
\includegraphics[scale=0.28]{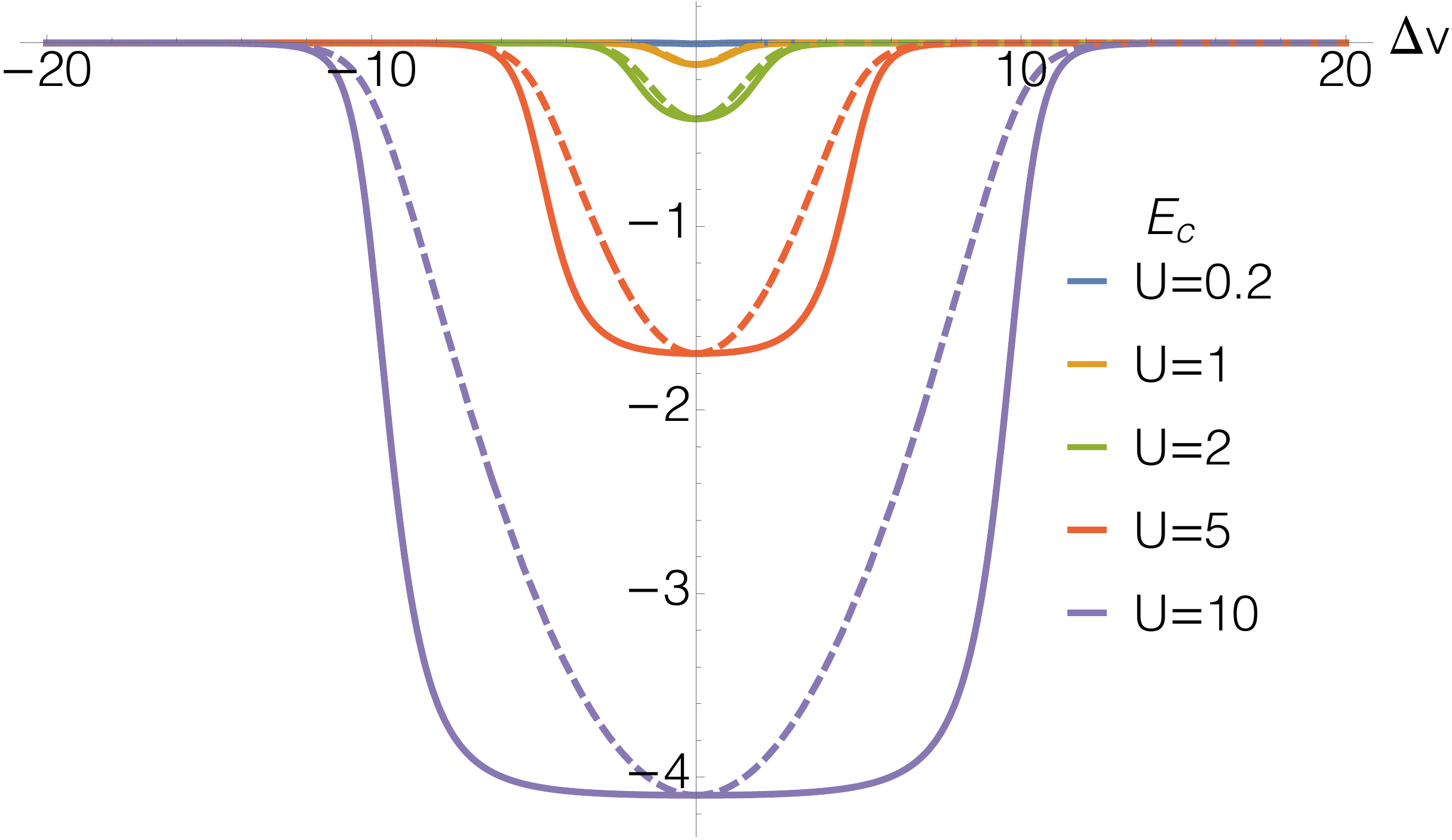}
\caption{Total correlation energies  $E_c^\hf$ (dashed) and $ E_c^\ks $ (solid) respectively Eqs.~\eqref{eq:HFEc} and~\eqref{eq:KSEc} for the Hubbard dimer as a function of $ \Delta v $, for $ U =0.2, 1, 2, 5, 10$.} 
\label{fig:Ecksvshf}
\end{figure}

Finally, we examine how $ E_c^\hf $ and $ E_c^\ks $ compare to one another. Both correlation energies account for the difference between the expectation value of the Hamiltonian operator in the GS and that in the single SD reference state, i.e.,
\begin{equation}
E_c^\text{SD} = E-\langle \Phi^\text{SD}| \hat{H}| \Phi^\text{SD} \rangle,
\end{equation}
with $ \text{SD}=\hf, \ks $. By virtue of definition~\eqref{eq:HFwf} and of the variational principle, 
it is immediate to see that $ E_c^\ks \leq E_c^\hf $. However, as demonstrated above, both $ U_c^\hf $ and $ T_c^\hf $ are less than or equal to their KS counterparts, meaning $ T_c^\hf + U_c^\hf  \leq T_c^\ks + U_c^\ks $. 
We conclude that it is the term $ V_c^\hf $ of Eq.~\eqref{eq:Vc}, shown in Fig.~\ref{fig:Vchf}, that cancels out a significant portion of the error residing in the other contributions ($T_c^\hf$ and $ U_c^\hf$).
Thus, adding up all the terms, we indeed retrieve the inequality $  E_c^\ks \leq  E_c^\hf $, holding for a given external potential, as seen in Fig.~\ref{fig:Ecksvshf}. 
%
Questions remain about how the results illustrated so far for the Hubbard dimer transfer to Coulomb quantum systems in the spatial continuum. Due to the profound difference in nature between the hopping operator and the quantum kinetic energy operator, it is hard to say whether there exist more realistic quantum systems where the HF kinetic energy could be higher than the interacting one. Typically, we expect the kinetic energy of particles interacting via an effective mean field to be lower than the one of electrons interacting coulombically. But we do not exclude that, at least outside of equilibrium geometries, it might be possible to find rather pathological examples of this unusual inverted relationship between the HF and true kinetic energy. That said, in Reference~\onlinecite{GriSchBae-JCP-97}, the authors study the individual contributions to the correlation energies within (R)HF and KS theories for the molecules Li$_2 $, N$_2 $ and F$_2 $ at equilibrium or larger bond distances. There, the case where $T_c^\hf < T_c^\ks$ (as in Fig.~\ref{fig:Tcksvshf}) is never encountered. In other words, in their cases the HF kinetic energy is typically lower than the KS one for a given external potential and, consequently, lower than the interacting one. However, they also observe that the HF kinetic energy is typically much more sensitive to the geometry than the KS one.

Concerning the remaining contributions, $ U_c^\text{SD} $ and $ V_c^\hf $, our results are of somewhat general validity: the HF state tends to ``overstabilize" the energy by relaxing the density, but the individual contributions to the energy are less in line with the exact ones than their KS counterparts. This is in agreement with what is observed in Reference~\onlinecite{GriSchBae-JCP-97}, namely that: $U_{H,c}^\hf$ and $V_c^\hf$ have similar orders of magnitude (with this latter being typically larger, up to a factor of four) and are opposite in sign, while the difference between $U_{c, ind}^\hf$ and $U_c^\ks$ is between one and three orders of magnitudes smaller (in the Hubbard dimer, as said, this difference is exactly zero). The only \textit{caveat} is that, in non-lattice systems, the HF density is typically more diffuse. This means $ V_c^\hf < 0$, as the HF density is less peaked around the nuclei where the nuclear field is more attractive, and that $ U_{H,c}^\hf >0 $, as the HF mean field repulsion is milder. In our model, a more diffuse density translates in a larger site occupation difference, i.e., $| \Delta n^\hf |\geq |\Delta n| $ (see Fig.~\ref{fig:NNinset}), resulting in those contributions having the reverse sign, i.e. $ V_c^\hf > 0$ and $ U_{c,H} <0 $.

\begin{figure}[ht]
\includegraphics[scale=0.28]{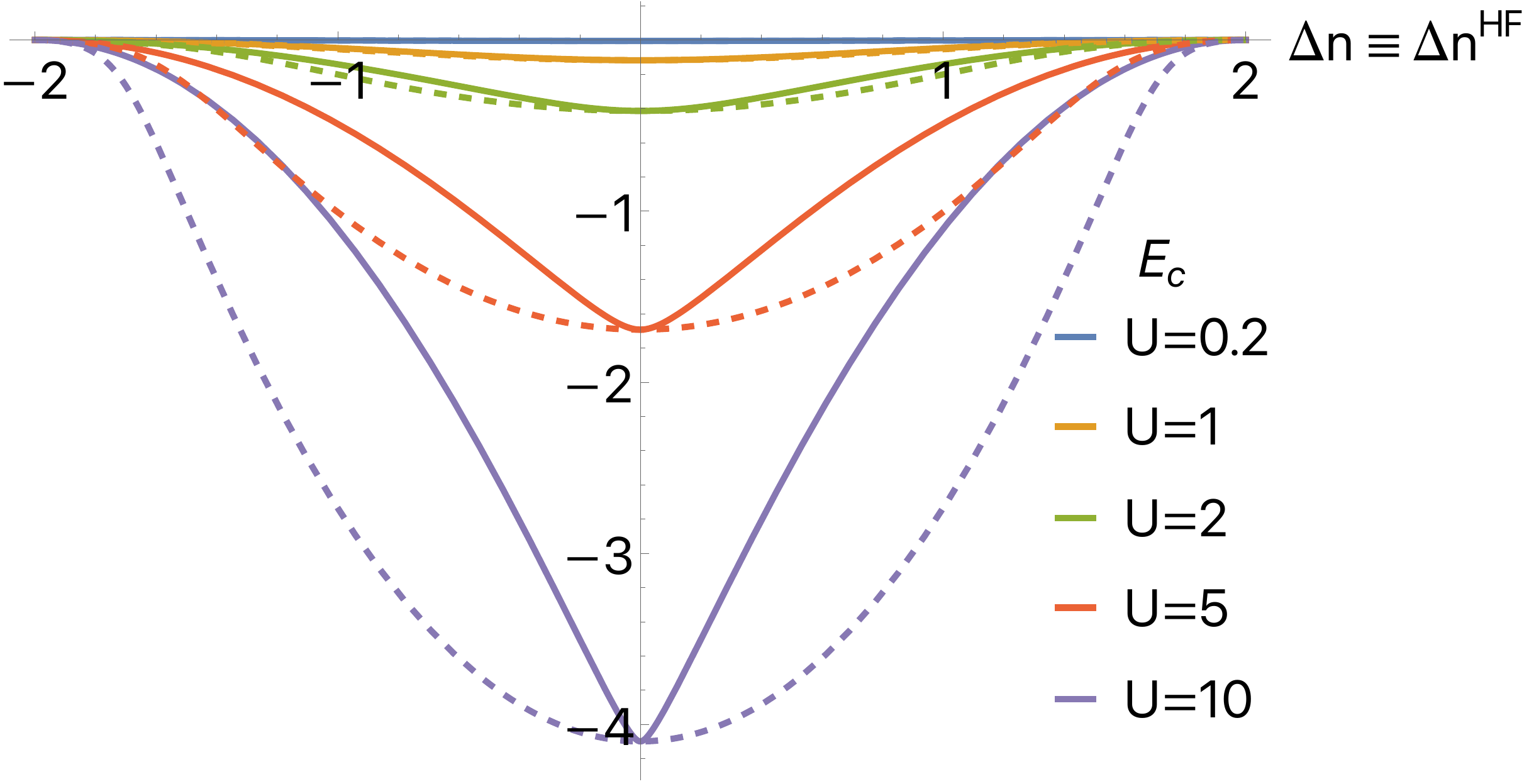}
\caption{Total correlation energies  $E_c^\hf$ (dashed) and $ E_c^\ks $ (solid) -- Eqs.~\eqref{eq:HFEc} and~\eqref{eq:KSEc}, respectively -- for the Hubbard dimer as a function of the site occupation $\Delta n^\hf$ and $\Delta n$ (set equal), for $ U =0.2, 1, 2, 5, 10$. } 
\label{fig:Ecvsn}
\end{figure}

We now want to consider a scenario which can virtually be realised only within the Hubbard dimer setting. Namely, we ask ourselves how the HF and the KS correlation energy functions compare to one another if we match the the two site occupation differences. In this case, as visible in Fig.~\ref{fig:Ecvsn}, the opposite inequality appears to hold, i.e.,
\begin{equation}
E_c^\ks (U, \Delta n )\big|_{\Delta n \equiv \Delta n^\hf} \geq  E_c^\hf (U, \Delta n^\hf).
\end{equation}  

Note that the function $E_c^\ks (U, \Delta n)$ is not known analytically and it has been obtained from numerical inversion.
On the other hand, the expression for $E_c^\hf$ can be found analytically and reads
\begin{widetext}
\begin{equation}\label{eq:EchfHD}
E_c^\hf (U,\,x)=\frac{1}{24} \left(-16 \, f(U,\,x) \sin \left(\frac{1}{6} \left(2 \cos ^{-1}\left(-\frac{U}{2}\frac{
g(U,\,x)}{f(U,\,x)^{3}}\right)+\pi \right)\right)+3 U x^2+4 U+\frac{48}{\sqrt{4-x^2}}\right), 
\end{equation}
\end{widetext}
with
\begin{eqnarray}
& &f(U,\,x)=\sqrt{3+U^2+3 x^2 \left(\frac{U}{2}+\frac{1}{\sqrt{4-x^2}}\right)^2},\notag \\
& &g(U,\,x)=\left(9+2 U^2-18 x^2 \left(\frac{U}{2}+\frac{1}{\sqrt{4-x^2}}\right)^2\right),\notag
\end{eqnarray}
and $x=\Delta n^\hf$.
Its small-$U$ expansion
gives us the M\o ller-Plesset perturbation~\cite{MolPle-PR-34} series coefficients
\begin{eqnarray}
E_c^\text{MP2}(U,\,x)& = &-\frac{1}{256} U^2 \left(4-x^2\right)^{5/2},\label{eq:Ecmp2HD}\\
 E_c^\text{MP3}(U,\,x)& = &-\frac{U^3 x^2 \left(x^2-4\right)^3}{2048}, \label{eq:Ecmp3HD}\\
 ... \notag
\end{eqnarray}
(where we have reported only the first two, as the coefficients grow in complexity). 

Comparison of Eqs.~\eqref{eq:Ecmp2HD} and~\eqref{eq:Ecmp3HD} with the G\"orling-Levy~\cite{GorLev-PRB-93, GorLev-PRA-94} series expansion coefficients, $ E_c^\text{GL2}$ and $E_c^\text{GL3}$, reported in Eqs.~(88) and~(89) of Reference~\onlinecite{CarFerSmiBur-JPCM-15}, shows that these coefficients are formally identical in the two perturbation treatments for the Hubbard dimer. The only difference is that, here, they are a function of the HF site occupation, $x=\Delta n^\hf$, whereas in the DFT case, they are functions of the interacting site occupation. (Note also that each of the terms in the energy expressions of Eq.~\eqref{eq:EchfHD}, \eqref{eq:Ecmp2HD} and \eqref{eq:Ecmp3HD} depends on the square of the site occupation difference, rather than on the site occupation difference itself.) 
Concerning the second-order coefficients, their formal equivalence is due to the aforementioned lack of exchange in this model. As for the equivalence between the third-order coefficients, it may be due simply to the lack of exchange, however similar investigations in other models are needed to clarify its influence.

To summarize, in this section, we have calculated the \emph{exact} total and partial correlation energies corresponding to the HF or the KS reference states, comparing the resulting pairs. In the following section, we shall focus on \emph{approximate} expressions for the correlation energy which can be used within both theories.
\section{Performance of the LB and SPL functionals for the Hubbard dimer}
We previously introduced the practice of adopting density functional approximations developed for the correlation energy in KS-DFT and using them with HF ingredients as a correction to the HF energy. In particular, functionals coming from the so-called adiabatic connection framework, ACMs, have been successfully used in this manner.~\cite{FabGorSeiDel-JCTC-16, VucGorDelFab-JPCL-18, GiaGorDelFab-JCP-18, DaaFabSalGorVuc-JPCL-21} As said, these formulas interpolate between the weak- and strong-interaction expansions of the adiabatic connection integrand, $W_\l^\sd$. This function(al) integrates to the desired correlation energy between the two extremes, zero and one, of the interaction strength parameter $\l$, i.e. $\int_0^1 W_\l^\sd \ud \l =E_c^\sd $ (with SD=HF, KS). 
A more detailed treatment of the MP adiabatic connection integrand, $W_\l^\hf$, for the Hubbard dimer is currently in preparation. In this context, we focus only on the performances of the adiabatic connection methods corresponding to the LB [Eq.~\eqref{eq:LB}] and the SPL [Eq.~\eqref{eq:SPL}] functionals. The validity of such approximations, in the Hubbard dimer setting, can be assessed without introducing any other source of errors, such as the ones coming from using approximate KS orbitals (e.g. PBE,~\cite{PerBurErn-PRL-96} PBE0,~\cite{pbe0_2} etc) or basis set expansions. These LB and SPL formulas require as ingredients the quantities $E_x$, $E_c^\text{PT2}$ and $W_\infty$.
The first one, as said, is exactly zero in the Hubbard dimer, thus we have $\tilde{W}_\infty\equiv W_\infty$ in this case.
The $E_c^\text{PT2}$ ingredient corresponds to Eq.~\eqref{eq:Ecmp2HD} for both references (HF and KS), as discussed in the previous section.

$W_\infty$ corresponds to the leading term, which is order $\l$, in the large-$\l$ expansion of $E_c^\sd(\l U, x)$:
\begin{equation}
\lim_{\l\to\infty}E_c^\sd(\l U, x) \sim \l W_\infty^\sd.
\end{equation}
Explicit expressions for $W_\infty^\sd$'s two different reference states read
\begin{equation}
 W_\infty^\hf(U, \,x) = \frac{U}{8} \left( x^2 -4 \right),
\end{equation}
and
\begin{equation}
W_\infty^\ks(U, \,x) = -\frac{U}{2}\left( 1-\Big| \frac{x}{2}\Big|\right)^2.
\end{equation}
Note that the latter expression has been already reported in Eq.~(56) of Reference~\onlinecite{CarFerSmiBur-JPCM-15} (as subsequently corrected in the Erratum~\cite{CarFerSmiBur-JPCM-16}).

\begin{figure}
\includegraphics[scale=0.28]{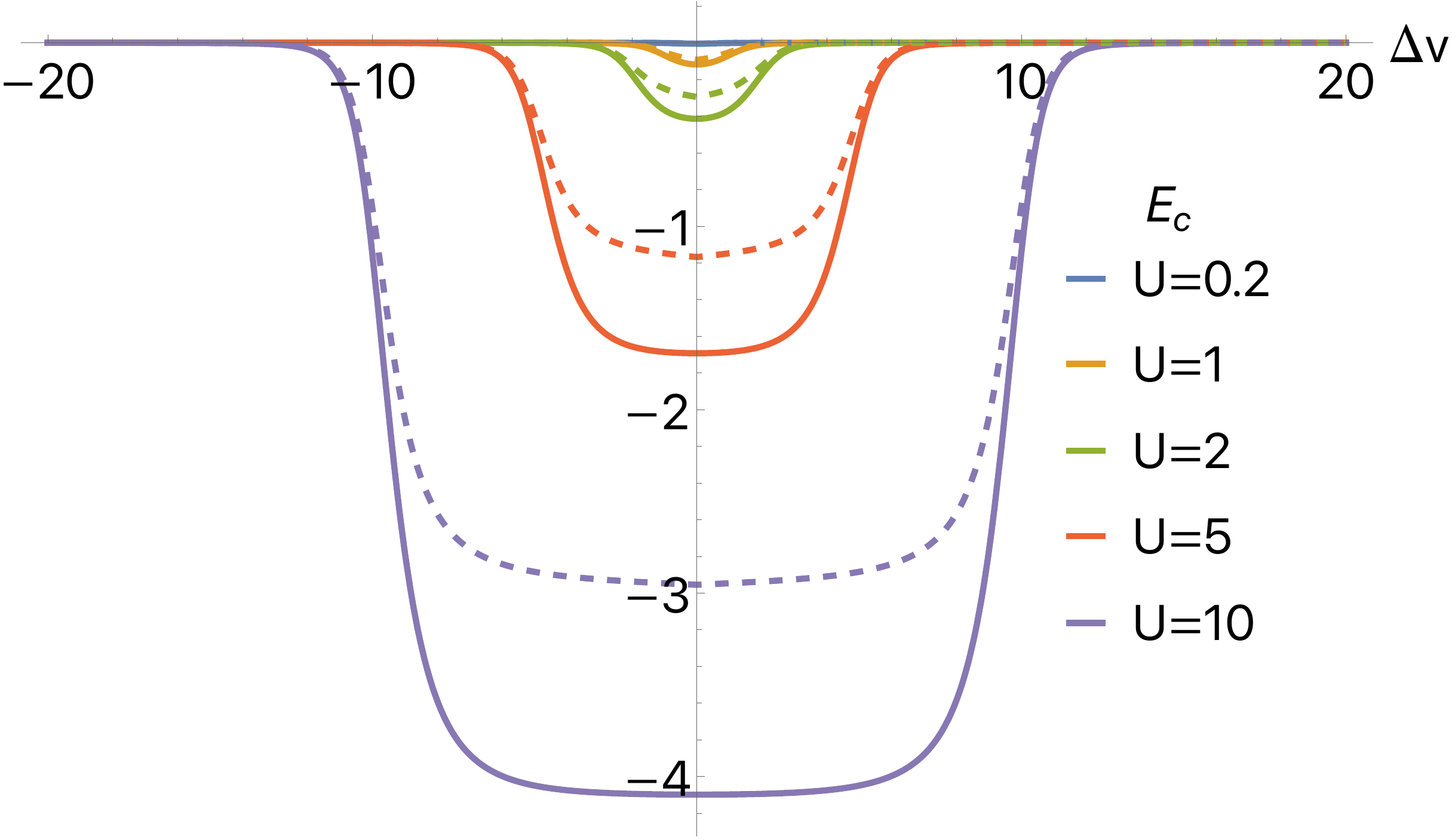}
\caption{Exact KS correlation energy $E_c^\ks$ (thick) and its approximation using the LB functional [Eq.~\eqref{eq:LB}] (dashed) for the Hubbard dimer with ingredients $W_0'=2\,E_c^\text{GL2}$, $W_\infty = W_\infty^\ks(U, \,x)$ and $x=\Delta n (\Delta v)$, for $ U =0.2, 1, 2, 5, 10$.} 
\label{fig:Eclbdft}
\end{figure}

\begin{figure}
\includegraphics[scale=0.28]{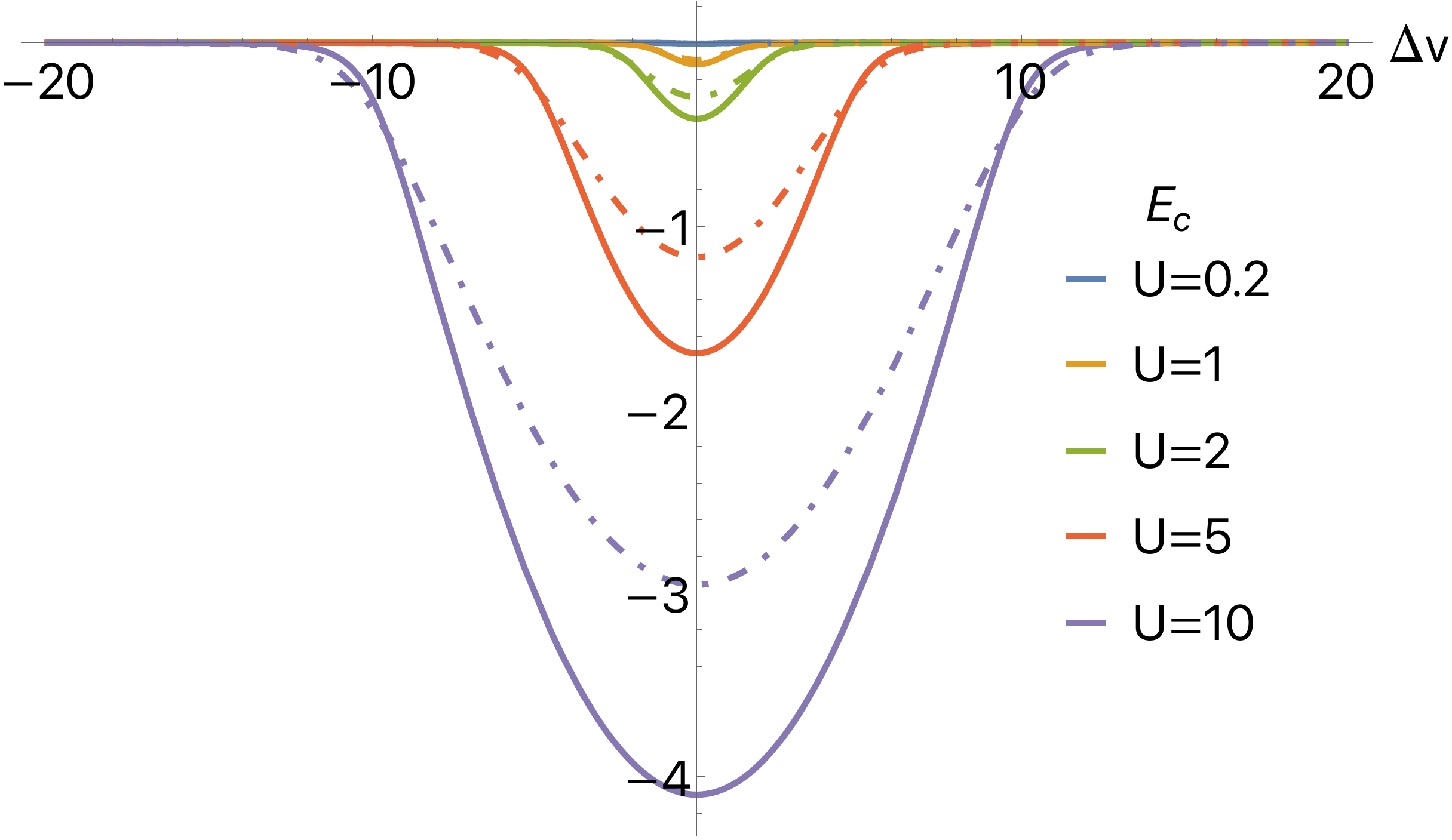}
\caption{Exact HF correlation energy $E_c^\hf$ (thick) and its approximation using the LB functional [Eq.~\eqref{eq:LB}] (dot-dashed) for the Hubbard dimer with ingredients $W_0'=2\,E_c^\text{MP2}$, $W_\infty = W_\infty^\hf(U, \,x)$ and $x=\Delta n^\hf (\Delta v)$, for $ U =0.2, 1, 2, 5, 10$.} 
\label{fig:Eclbhf}
\end{figure}

\begin{figure}
\includegraphics[scale=0.28]{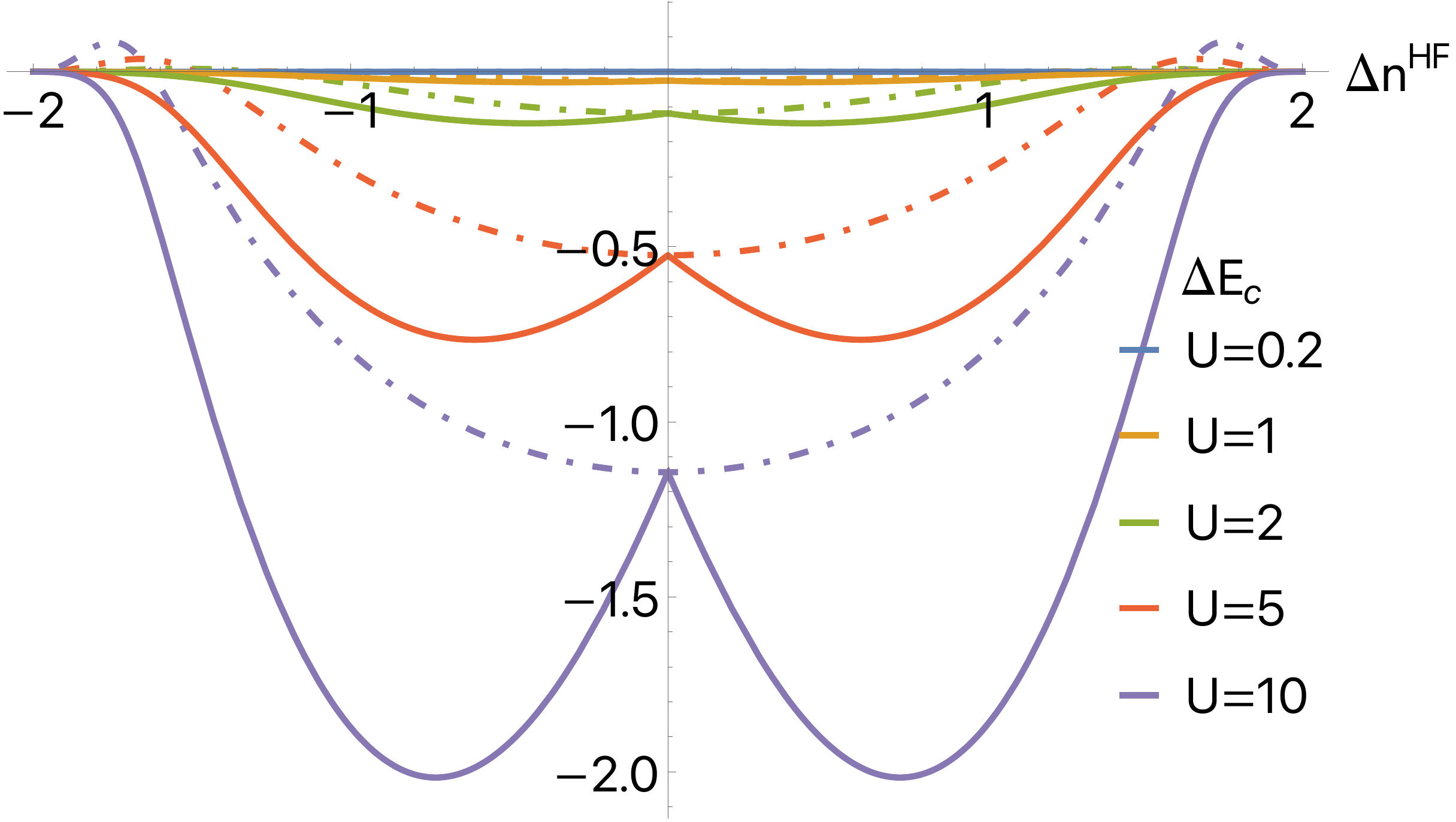}
\caption{Difference $\Delta E_c^\text{LB}=\left(E_c -E_c^\text{LB} \right)$ for HF reference state (dot-dashed) and for $E_c^\text{LB}$ with mixed ingredients, namely $W_0'=2\, E_c^\text{MP2}$, $W_\infty = W_\infty^\ks(U, \,x)$ and $x=\Delta n^\hf$ as a correction to the HF energy, at various $U$, as a function of $\Delta n^\hf$.} 
\label{fig:deltaEclbmixed}
\end{figure}

In Fig.~\ref{fig:Eclbdft}, we compare how well the LB approximation works for the Hubbard dimer in the context of KS-DFT.
In Fig.~\ref{fig:Eclbhf}, we report instead the performance of the LB functional used with HF ingredients as a correction to the traditional correlation energy, $E_c^\hf$.
Finally, in Fig.~\ref{fig:deltaEclbmixed}, we plot the difference
$\Delta E_c = \left(E_c -E_c^\text{LB} \right)$ for each method. As is visible, for a large portion of the parameter space, the LB approximation for the Hubbard dimer works better for the HF reference state and as a correction to the traditional correlation energy than for the KS ones. In fact, the only region where the LB approximation works better for the KS correlation energy corresponds to weakly correlated systems, where the external potential difference dominates over the repulsion term.  We also note that, just as for the HF kinetic energy (see Fig.~\ref{fig:Tcksvshf}), there is a particular combination of $U$ and $\Delta v$ for which the LB approximation yields the exact HF correlation energy.

\begin{figure}
\includegraphics[scale=0.28]{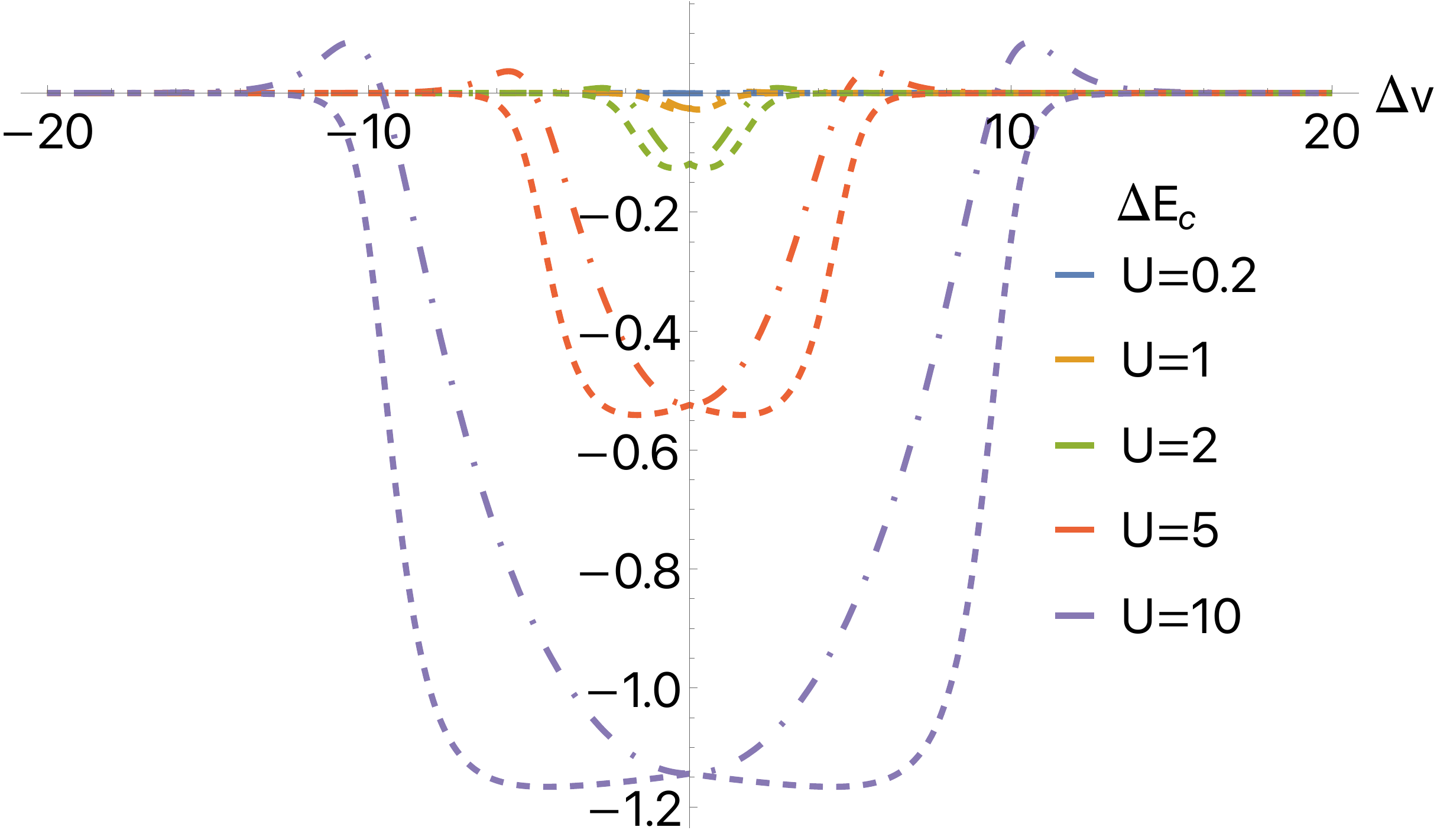}
\caption{Difference $\Delta E_c^\text{LB}=\left( E_c -E_c^\text{LB}\right)$ for HF reference state (dot-dashed) and for the KS reference state (dashed) at various $U$, as a function of $\Delta v$.} 
\label{fig:deltaEclbdfthf}
\end{figure}

As a further point, we propose to investigate the performance of the LB functional adopting mixed ingredients: the $W_\infty$ associated to the KS-DFT correlation energy, but with the HF site occupation difference as input.
This may seem quite an arbitrary choice. However, it is precisely the way in which said ACMs have mostly been used, for a very pragmatic reason. Whereas $W_\infty^\ks$ has been known for quite a long time,~\cite{Sei-PRA-99} and an excellent approximation to it in the form of a gradient expansion has been developed since,~\cite{SeiPerKur-PRA-00} $W_\infty^\hf$ has been introduced only recently,~\cite{SeiGiaVucFabGor-JCP-18} and gradient expansion approximations to it have just been devised.~\cite{DaaKooGroSeiGor-JCTC-22} The mixed LB functional thus obtained is used as a correction to the HF energy. Its performance for the Hubbard dimer is shown in Fig.~\ref{fig:deltaEclbmixed}, contrasted with the internally-consistent strategy already discussed, as a function of the HF site occupation.
For most of the site occupation domain, this mixed-ingredient combination is quite inaccurate, greatly worsening in performance. There is only a small region where the mixed-ingredient combination yields better estimates of the HF correlation energy. This region corresponds to the outer edges of the domain of $\Delta n^\hf$, i.e. the weakly-correlated cases where $|\Delta n^\hf|$ approaches two. 

Therefore, in the Hubbard dimer setting, it is clear that the LB functional (as well as the SPL functional, see discussion below) works better when the appropriate strong-interaction ingredient for the HF reference, i.e. $W_\infty^\hf$ is adopted, rather than $W_\infty^\ks$, for high- and intermediate-correlation regimes. This is somewhat reassuring, as it shows that these adiabatic connection methods work as intended, giving better results when consistent ingredients are used and not benefiting from an error cancellation between the KS ingredient $W_\infty^\ks$ and the HF site-occupation input. 

The trends observed for the SPL functional across the Hubbard dimer parameter space were qualitatively equivalent to those observed for the LB functional, though the SPL estimate of the correlation energy appears to be larger than the LB one everywhere, for both the KS and the HF references cases.
Note that, in the KS case, both LB and SPL functionals appear to bound the exact correlation energies from above. This, in turn, means that the SPL correlation energy error is everywhere larger than the LB one, i.e. $|\Delta E_c^\text{KS,SPL}|>|\Delta E_c^\text{KS, LB}|$ with $\Delta E_c^\text{KS, ACM} = E_c^\ks - E_c^\text{KS, ACM}$ and ACM$=$LB, SPL.
As an example, the maximum error for $U=10$ is 1.14 E$_h$ for the LB functional and 1.42 E$_h$ for the SPL one.

As for the case of the HF reference, both functionals have a turning point around the value $|\Delta v| \approx U$. In the strong correlation regime, when $|\Delta v| < U$, they underestimate (in magnitude) the exact correlation energy. Past the turning point, when $|\Delta v| > U$, they `overshoot' it. This, in turn, means that the SPL correlation energy error is larger than the LB one, $|\Delta  E_c^\text{HF,SPL}|>|\Delta E_c^\text{HF, LB}|$ in the more strongly correlated cases where $|\Delta v| < U$. In the weakly correlated range of the parameter space (i.e., where $|\Delta v| > U$), we have instead $|\Delta  E_c^\text{HF,SPL}|<|\Delta E_c^\text{HF, LB}|$. Finally, the use of mixed ingredients worsens the performance of the SPL functional in a manner essentially analogous to that observed for the LB case in Fig.~\ref{fig:deltaEclbmixed}. The fact that the two different functionals show such a close similarity of trends across the Hubbard dimer parameter space may indicate that the common rationale underpinning both functionals largely determines their performances, despite their differences. A detailed account of the results of the SPL functional, similar to those shown in Figs.~\ref{fig:Eclbdft},~\ref{fig:Eclbhf}, \ref{fig:deltaEclbdfthf} and~\ref{fig:deltaEclbmixed} for the LB one, can be found in the supporting notebook contained in the Supplemental Information.

\section{Conclusions}

We have provided an analytical comparison between HF and KS-DFT methods for the Hubbard dimer model. One of the most striking findings within this model is that the indirect interaction energies for the two methods, $U_{c,ind}^\text{SD}$, are exactly the same at a given external potential. In line with Ref.~\onlinecite{GriSchBae-JCP-97}, our results show that the HF solution can ``overstabilize” the energy through the external potential by relaxing the density (site-occupation). However, the separate contributions to the energy typically deviate more from the corresponding interacting ones than their KS counterparts (see Figs.~\ref{fig:Tcksvshf} and~\ref{fig:Ucksvshf}). A notable exception is our demonstration of the change in sign in the HF kinetic correlation, which we understand to be a novel finding and contrary to intuitive predictions of its behavior. Furthermore, as the mapping between external potential and HF site-occupation is analytically invertible, unlike the interacting case, it is possible to obtain the exact correlation energy that corrects the HF approximation as a function of the HF density in a pure site-occupation function theory (SOFT) spirit [see Eq.~\eqref{eq:EchfHD}]. 


On the subject of adiabatic connection methods, we have assessed the performances of the LB and SPL functionals, finding that, for the more strongly-interacting cases, they work better as an approximation to $E_c^\hf$, rather than as approximations to $E_c^\ks$, as originally intended. 
Note that, in our assessment, we were able to adopt the exact strong-interaction ingredient corresponding to the HF reference ($W_\infty^\hf$). This is not ordinary. In fact, several works pioneering the application of ACMs as a correction to the HF energy used 
(a model for) the DFT strong-interaction ingredient, $W_\infty^\ks$, with the HF density.~\cite{FabGorSeiDel-JCTC-16, VucGorDelFab-JPCL-18, GiaGorDelFab-JCP-18}
The only exception is a recent work in which an empirical model for $W_\infty^\hf$ is adopted.~\cite{DaaFabSalGorVuc-JPCL-21} In turn, as shown in Fig.~\ref{fig:deltaEclbmixed}, the use of $W_\infty^\ks$ with the HF density greatly worsens the performances of the ACMs considered, supporting the view that an improvement of their performances on real molecules might follow from using the approximation for the HF strong-interaction ingredient that has recently become available.~\cite{DaaKooGroSeiGor-JCTC-22}

Given that the exchange energy term is absent from the Hubbard model setting, we have limited our analyses to the correlation part of the energy. Nonetheless, our conclusions on how the HF and KS methods compare from a formal point of view should not vary much by an inclusion of the exchange energy. In fact, generally, the exact exchange energy in the two references is expected to differ only slightly.~\cite{GriSchBae-JCP-97} As for how the examined ACMs perform according to which reference is used (HF or KS), the inclusion of the exchange energy term does not affect our conclusions because these methods recover full exact exchange (in other words, the exchange energy term is merely a constant shift). The situation in which one calculates the self-consistent density coming from the chosen ACM applied within the KS-DFT framework would be different, however. This would give an approximation for the KS quantities input in the correlation energy functional that would reflect on its outcome, as well as on the exchange energy. Since the HF framework for the ACMs demands that the correlation energy is added as a post-self-consistent-field correction using HF orbitals as input, to compare the performances of these interpolations across the two methods we have considered only their application on exact KS quantities and not their self-consistent-field solution. Nonetheless, this is an aspect to keep in mind since the way in which these ACMs can be used in actual KS-DFT calculations requires either an underlying density functional model to determine approximate KS orbitals or an SCF implementation. An investigation in the Hubbard dimer setting, especially in light of the computational cost for an SCF implementation on real molecules, could be instructive. In fact, an SCF implementation of these ACMs has been carried out only very recently and tested for few simple chemical species (Ne, CO and H$_2$).~\cite{SmiSalGorFab-arxiv-21} Follow-up work, in which we present a detailed analysis of the adiabatic connection integrand corresponding to the HF reference [Eq~\eqref{eq:Wlmp}] ~\cite{SeiGiaVucFabGor-JCP-18} for the Hubbard dimer, is currently in progress.

\section{Acknowledgments} This work is supported by the U.S. Department of Energy, National Nuclear Security Administration, Minority Serving Institution Partnership Program, under Award DE-NA0003866. We acknowledge all indigenous peoples local to the site of University of California, Merced, including the Yokuts and Miwuk. We embrace their continued connection to this region and thank them for allowing us to live, work, learn, and collaborate on their traditional homeland.

\nocite{*}
\bibliography{aipsamp}

\providecommand{\noopsort}[1]{}\providecommand{\singleletter}[1]{#1}%
\begin{thebibliography}{45}%
\makeatletter
\providecommand \@ifxundefined [1]{%
 \@ifx{#1\undefined}
}%
\providecommand \@ifnum [1]{%
 \ifnum #1\expandafter \@firstoftwo
 \else \expandafter \@secondoftwo
 \fi
}%
\providecommand \@ifx [1]{%
 \ifx #1\expandafter \@firstoftwo
 \else \expandafter \@secondoftwo
 \fi
}%
\providecommand \natexlab [1]{#1}%
\providecommand \enquote  [1]{``#1''}%
\providecommand \bibnamefont  [1]{#1}%
\providecommand \bibfnamefont [1]{#1}%
\providecommand \citenamefont [1]{#1}%
\providecommand \href@noop [0]{\@secondoftwo}%
\providecommand \href [0]{\begingroup \@sanitize@url \@href}%
\providecommand \@href[1]{\@@startlink{#1}\@@href}%
\providecommand \@@href[1]{\endgroup#1\@@endlink}%
\providecommand \@sanitize@url [0]{\catcode `\\12\catcode `\$12\catcode
  `\&12\catcode `\#12\catcode `\^12\catcode `\_12\catcode `\%12\relax}%
\providecommand \@@startlink[1]{}%
\providecommand \@@endlink[0]{}%
\providecommand \url  [0]{\begingroup\@sanitize@url \@url }%
\providecommand \@url [1]{\endgroup\@href {#1}{\urlprefix }}%
\providecommand \urlprefix  [0]{URL }%
\providecommand \Eprint [0]{\href }%
\providecommand \doibase [0]{http://dx.doi.org/}%
\providecommand \selectlanguage [0]{\@gobble}%
\providecommand \bibinfo  [0]{\@secondoftwo}%
\providecommand \bibfield  [0]{\@secondoftwo}%
\providecommand \translation [1]{[#1]}%
\providecommand \BibitemOpen [0]{}%
\providecommand \bibitemStop [0]{}%
\providecommand \bibitemNoStop [0]{.\EOS\space}%
\providecommand \EOS [0]{\spacefactor3000\relax}%
\providecommand \BibitemShut  [1]{\csname bibitem#1\endcsname}%
\let\auto@bib@innerbib\@empty
\bibitem [{\citenamefont {Sharkas}\ \emph {et~al.}(2011)\citenamefont
  {Sharkas}, \citenamefont {Toulouse},\ and\ \citenamefont
  {Savin}}]{ShaTouSav-JCP-11}%
  \BibitemOpen
  \bibfield  {author} {\bibinfo {author} {\bibfnamefont {K.}~\bibnamefont
  {Sharkas}}, \bibinfo {author} {\bibfnamefont {J.}~\bibnamefont {Toulouse}}, \
  and\ \bibinfo {author} {\bibfnamefont {A.}~\bibnamefont {Savin}},\
  }\href@noop {} {\bibfield  {journal} {\bibinfo  {journal} {J. Chem. Phys.}\
  }\textbf {\bibinfo {volume} {134}},\ \bibinfo {pages} {064113} (\bibinfo
  {year} {2011})}\BibitemShut {NoStop}%
\bibitem [{\citenamefont {Ghosh}\ \emph {et~al.}(2018)\citenamefont {Ghosh},
  \citenamefont {Verma}, \citenamefont {Cramer}, \citenamefont {Gagliardi},\
  and\ \citenamefont {Truhlar}}]{GhoVerCraGagTru-CR-18}%
  \BibitemOpen
  \bibfield  {author} {\bibinfo {author} {\bibfnamefont {S.}~\bibnamefont
  {Ghosh}}, \bibinfo {author} {\bibfnamefont {P.}~\bibnamefont {Verma}},
  \bibinfo {author} {\bibfnamefont {C.~J.}\ \bibnamefont {Cramer}}, \bibinfo
  {author} {\bibfnamefont {L.}~\bibnamefont {Gagliardi}}, \ and\ \bibinfo
  {author} {\bibfnamefont {D.~G.}\ \bibnamefont {Truhlar}},\ }\href@noop {}
  {\bibfield  {journal} {\bibinfo  {journal} {Chem. Rev.}\ }\textbf {\bibinfo
  {volume} {118}},\ \bibinfo {pages} {7249} (\bibinfo {year}
  {2018})}\BibitemShut {NoStop}%
\bibitem [{\citenamefont {Vuckovic}\ \emph {et~al.}(2019)\citenamefont
  {Vuckovic}, \citenamefont {Song}, \citenamefont {Kozlowski}, \citenamefont
  {Sim},\ and\ \citenamefont {Burke}}]{VucSonKozSimBur-JCTC-19}%
  \BibitemOpen
  \bibfield  {author} {\bibinfo {author} {\bibfnamefont {S.}~\bibnamefont
  {Vuckovic}}, \bibinfo {author} {\bibfnamefont {S.}~\bibnamefont {Song}},
  \bibinfo {author} {\bibfnamefont {J.}~\bibnamefont {Kozlowski}}, \bibinfo
  {author} {\bibfnamefont {E.}~\bibnamefont {Sim}}, \ and\ \bibinfo {author}
  {\bibfnamefont {K.}~\bibnamefont {Burke}},\ }\href@noop {} {\bibfield
  {journal} {\bibinfo  {journal} {J. Chem. Theory. Comput.}\ } (\bibinfo {year}
  {2019})}\BibitemShut {NoStop}%
\bibitem [{\citenamefont {Gritsenko}\ \emph {et~al.}(1997)\citenamefont
  {Gritsenko}, \citenamefont {Schipper},\ and\ \citenamefont
  {Baerends}}]{GriSchBae-JCP-97}%
  \BibitemOpen
  \bibfield  {author} {\bibinfo {author} {\bibfnamefont {O.}~\bibnamefont
  {Gritsenko}}, \bibinfo {author} {\bibfnamefont {P.}~\bibnamefont {Schipper}},
  \ and\ \bibinfo {author} {\bibfnamefont {E.}~\bibnamefont {Baerends}},\
  }\href@noop {} {\bibfield  {journal} {\bibinfo  {journal} {J. Chem. Phys.}\
  }\textbf {\bibinfo {volume} {107}},\ \bibinfo {pages} {5007} (\bibinfo {year}
  {1997})}\BibitemShut {NoStop}%
\bibitem [{\citenamefont {Hubbard}(1963)}]{Hub-PRSL-63}%
  \BibitemOpen
  \bibfield  {author} {\bibinfo {author} {\bibfnamefont {J.}~\bibnamefont
  {Hubbard}},\ }\href@noop {} {\bibfield  {journal} {\bibinfo  {journal}
  {Proceedings of the Royal Society of London. Series A. Mathematical and
  Physical Sciences}\ }\textbf {\bibinfo {volume} {276}},\ \bibinfo {pages}
  {238} (\bibinfo {year} {1963})}\BibitemShut {NoStop}%
\bibitem [{\citenamefont {Lieb}\ and\ \citenamefont {Wu}(1968)}]{LieWu-PRL-68}%
  \BibitemOpen
  \bibfield  {author} {\bibinfo {author} {\bibfnamefont {E.~H.}\ \bibnamefont
  {Lieb}}\ and\ \bibinfo {author} {\bibfnamefont {F.}~\bibnamefont {Wu}},\
  }\href@noop {} {\bibfield  {journal} {\bibinfo  {journal} {Physical Review
  Letters}\ }\textbf {\bibinfo {volume} {20}},\ \bibinfo {pages} {1445}
  (\bibinfo {year} {1968})}\BibitemShut {NoStop}%
\bibitem [{\citenamefont {Montorsi}(1992)}]{Mon-book-92}%
  \BibitemOpen
  \bibfield  {author} {\bibinfo {author} {\bibfnamefont {A.}~\bibnamefont
  {Montorsi}},\ }\href@noop {} {\emph {\bibinfo {title} {The Hubbard Model: A
  Reprint Volume}}}\ (\bibinfo  {publisher} {World Scientific},\ \bibinfo
  {year} {1992})\BibitemShut {NoStop}%
\bibitem [{\citenamefont {Theophilou}\ \emph {et~al.}(2018)\citenamefont
  {Theophilou}, \citenamefont {Buchholz}, \citenamefont {Eich}, \citenamefont
  {Ruggenthaler},\ and\ \citenamefont {Rubio}}]{TheBucEicRugRub-JCTC-18}%
  \BibitemOpen
  \bibfield  {author} {\bibinfo {author} {\bibfnamefont {I.}~\bibnamefont
  {Theophilou}}, \bibinfo {author} {\bibfnamefont {F.}~\bibnamefont
  {Buchholz}}, \bibinfo {author} {\bibfnamefont {F.}~\bibnamefont {Eich}},
  \bibinfo {author} {\bibfnamefont {M.}~\bibnamefont {Ruggenthaler}}, \ and\
  \bibinfo {author} {\bibfnamefont {A.}~\bibnamefont {Rubio}},\ }\href@noop {}
  {\bibfield  {journal} {\bibinfo  {journal} {J. Chem. Theory Comput.}\
  }\textbf {\bibinfo {volume} {14}},\ \bibinfo {pages} {4072} (\bibinfo {year}
  {2018})}\BibitemShut {NoStop}%
\bibitem [{\citenamefont {Lacombe}\ and\ \citenamefont
  {Maitra}(2020)}]{LacMar-PRL-20}%
  \BibitemOpen
  \bibfield  {author} {\bibinfo {author} {\bibfnamefont {L.}~\bibnamefont
  {Lacombe}}\ and\ \bibinfo {author} {\bibfnamefont {N.~T.}\ \bibnamefont
  {Maitra}},\ }\href@noop {} {\bibfield  {journal} {\bibinfo  {journal}
  {Physical Review Letters}\ }\textbf {\bibinfo {volume} {124}},\ \bibinfo
  {pages} {206401} (\bibinfo {year} {2020})}\BibitemShut {NoStop}%
\bibitem [{\citenamefont {Marie}\ \emph {et~al.}(2021)\citenamefont {Marie},
  \citenamefont {Burton},\ and\ \citenamefont {Loos}}]{MarBurLoo-JPCM-21}%
  \BibitemOpen
  \bibfield  {author} {\bibinfo {author} {\bibfnamefont {A.}~\bibnamefont
  {Marie}}, \bibinfo {author} {\bibfnamefont {H.~G.}\ \bibnamefont {Burton}}, \
  and\ \bibinfo {author} {\bibfnamefont {P.-F.}\ \bibnamefont {Loos}},\
  }\href@noop {} {\bibfield  {journal} {\bibinfo  {journal} {Journal of
  Physics: Condensed Matter}\ } (\bibinfo {year} {2021})}\BibitemShut {NoStop}%
\bibitem [{\citenamefont {Carrascal}\ \emph {et~al.}(2015)\citenamefont
  {Carrascal}, \citenamefont {Ferrer}, \citenamefont {Smith},\ and\
  \citenamefont {Burke}}]{CarFerSmiBur-JPCM-15}%
  \BibitemOpen
  \bibfield  {author} {\bibinfo {author} {\bibfnamefont {D.}~\bibnamefont
  {Carrascal}}, \bibinfo {author} {\bibfnamefont {J.}~\bibnamefont {Ferrer}},
  \bibinfo {author} {\bibfnamefont {J.~C.}\ \bibnamefont {Smith}}, \ and\
  \bibinfo {author} {\bibfnamefont {K.}~\bibnamefont {Burke}},\ }\href@noop {}
  {\bibfield  {journal} {\bibinfo  {journal} {Journal of Physics: Condensed
  Matter}\ }\textbf {\bibinfo {volume} {27}},\ \bibinfo {pages} {393001}
  (\bibinfo {year} {2015})}\BibitemShut {NoStop}%
\bibitem [{\citenamefont {Cohen}\ and\ \citenamefont
  {Mori-S{\'a}nchez}(2016)}]{CohMor-PRA-16}%
  \BibitemOpen
  \bibfield  {author} {\bibinfo {author} {\bibfnamefont {A.~J.}\ \bibnamefont
  {Cohen}}\ and\ \bibinfo {author} {\bibfnamefont {P.}~\bibnamefont
  {Mori-S{\'a}nchez}},\ }\href@noop {} {\bibfield  {journal} {\bibinfo
  {journal} {Physical Review A}\ }\textbf {\bibinfo {volume} {93}},\ \bibinfo
  {pages} {042511} (\bibinfo {year} {2016})}\BibitemShut {NoStop}%
\bibitem [{\citenamefont {Ying}\ \emph {et~al.}(2016)\citenamefont {Ying},
  \citenamefont {Brosco}, \citenamefont {Lopez}, \citenamefont {Varsano},
  \citenamefont {Gori-Giorgi},\ and\ \citenamefont
  {Lorenzana}}]{YinBroLor-PRB-16}%
  \BibitemOpen
  \bibfield  {author} {\bibinfo {author} {\bibfnamefont {Z.-J.}\ \bibnamefont
  {Ying}}, \bibinfo {author} {\bibfnamefont {V.}~\bibnamefont {Brosco}},
  \bibinfo {author} {\bibfnamefont {G.~M.}\ \bibnamefont {Lopez}}, \bibinfo
  {author} {\bibfnamefont {D.}~\bibnamefont {Varsano}}, \bibinfo {author}
  {\bibfnamefont {P.}~\bibnamefont {Gori-Giorgi}}, \ and\ \bibinfo {author}
  {\bibfnamefont {J.}~\bibnamefont {Lorenzana}},\ }\href@noop {} {\bibfield
  {journal} {\bibinfo  {journal} {Physical Review B}\ }\textbf {\bibinfo
  {volume} {94}},\ \bibinfo {pages} {075154} (\bibinfo {year}
  {2016})}\BibitemShut {NoStop}%
\bibitem [{\citenamefont {Carrascal}\ \emph {et~al.}(2018)\citenamefont
  {Carrascal}, \citenamefont {Ferrer}, \citenamefont {Maitra},\ and\
  \citenamefont {Burke}}]{CarFerMaiBur-EPJB-18}%
  \BibitemOpen
  \bibfield  {author} {\bibinfo {author} {\bibfnamefont {D.~J.}\ \bibnamefont
  {Carrascal}}, \bibinfo {author} {\bibfnamefont {J.}~\bibnamefont {Ferrer}},
  \bibinfo {author} {\bibfnamefont {N.}~\bibnamefont {Maitra}}, \ and\ \bibinfo
  {author} {\bibfnamefont {K.}~\bibnamefont {Burke}},\ }\href@noop {}
  {\bibfield  {journal} {\bibinfo  {journal} {The European Physical Journal B}\
  }\textbf {\bibinfo {volume} {91}},\ \bibinfo {pages} {1} (\bibinfo {year}
  {2018})}\BibitemShut {NoStop}%
\bibitem [{\citenamefont {Senjean}\ \emph {et~al.}(2017)\citenamefont
  {Senjean}, \citenamefont {Tsuchiizu}, \citenamefont {Robert},\ and\
  \citenamefont {Fromager}}]{SenTsuRobFro-MP-17}%
  \BibitemOpen
  \bibfield  {author} {\bibinfo {author} {\bibfnamefont {B.}~\bibnamefont
  {Senjean}}, \bibinfo {author} {\bibfnamefont {M.}~\bibnamefont {Tsuchiizu}},
  \bibinfo {author} {\bibfnamefont {V.}~\bibnamefont {Robert}}, \ and\ \bibinfo
  {author} {\bibfnamefont {E.}~\bibnamefont {Fromager}},\ }\href@noop {}
  {\bibfield  {journal} {\bibinfo  {journal} {Molecular Physics}\ }\textbf
  {\bibinfo {volume} {115}},\ \bibinfo {pages} {48} (\bibinfo {year}
  {2017})}\BibitemShut {NoStop}%
\bibitem [{\citenamefont {Deur}\ \emph {et~al.}(2017)\citenamefont {Deur},
  \citenamefont {Mazouin},\ and\ \citenamefont {Fromager}}]{DeuMazFro-PRB-17}%
  \BibitemOpen
  \bibfield  {author} {\bibinfo {author} {\bibfnamefont {K.}~\bibnamefont
  {Deur}}, \bibinfo {author} {\bibfnamefont {L.}~\bibnamefont {Mazouin}}, \
  and\ \bibinfo {author} {\bibfnamefont {E.}~\bibnamefont {Fromager}},\
  }\href@noop {} {\bibfield  {journal} {\bibinfo  {journal} {Physical Review
  B}\ }\textbf {\bibinfo {volume} {95}},\ \bibinfo {pages} {035120} (\bibinfo
  {year} {2017})}\BibitemShut {NoStop}%
\bibitem [{\citenamefont {Smith}\ \emph {et~al.}(2016)\citenamefont {Smith},
  \citenamefont {Pribram-Jones},\ and\ \citenamefont
  {Burke}}]{SmiPriBur-PRB-16}%
  \BibitemOpen
  \bibfield  {author} {\bibinfo {author} {\bibfnamefont {J.~C.}\ \bibnamefont
  {Smith}}, \bibinfo {author} {\bibfnamefont {A.}~\bibnamefont
  {Pribram-Jones}}, \ and\ \bibinfo {author} {\bibfnamefont {K.}~\bibnamefont
  {Burke}},\ }\href@noop {} {\bibfield  {journal} {\bibinfo  {journal}
  {Physical Review B}\ }\textbf {\bibinfo {volume} {93}},\ \bibinfo {pages}
  {245131} (\bibinfo {year} {2016})}\BibitemShut {NoStop}%
\bibitem [{\citenamefont {Schipper}\ \emph {et~al.}(1998)\citenamefont
  {Schipper}, \citenamefont {Gritsenko},\ and\ \citenamefont
  {J.Baerends}}]{SchGriBae-TCA-98}%
  \BibitemOpen
  \bibfield  {author} {\bibinfo {author} {\bibfnamefont {P.~R.~T.}\
  \bibnamefont {Schipper}}, \bibinfo {author} {\bibfnamefont {O.~V.}\
  \bibnamefont {Gritsenko}}, \ and\ \bibinfo {author} {\bibfnamefont
  {E.}~\bibnamefont {J.Baerends}},\ }\href@noop {} {\bibfield  {journal}
  {\bibinfo  {journal} {Theor. Chim. Acc.}\ }\textbf {\bibinfo {volume} {99}},\
  \bibinfo {pages} {329} (\bibinfo {year} {1998})}\BibitemShut {NoStop}%
\bibitem [{\citenamefont {van Leeuwen}(2003)}]{Lee-AQC-03}%
  \BibitemOpen
  \bibfield  {author} {\bibinfo {author} {\bibfnamefont {R.}~\bibnamefont {van
  Leeuwen}},\ }\href@noop {} {\bibfield  {journal} {\bibinfo  {journal} {Adv.
  Quantum Chem.}\ }\textbf {\bibinfo {volume} {43}},\ \bibinfo {pages} {24}
  (\bibinfo {year} {2003})}\BibitemShut {NoStop}%
\bibitem [{\citenamefont {Giesbertz}\ and\ \citenamefont
  {Baerends}(2010)}]{GieBae-JCP-10}%
  \BibitemOpen
  \bibfield  {author} {\bibinfo {author} {\bibfnamefont {K.}~\bibnamefont
  {Giesbertz}}\ and\ \bibinfo {author} {\bibfnamefont {E.}~\bibnamefont
  {Baerends}},\ }\href@noop {} {\bibfield  {journal} {\bibinfo  {journal} {J.
  Chem. Phys.}\ }\textbf {\bibinfo {volume} {132}},\ \bibinfo {pages} {194108}
  (\bibinfo {year} {2010})}\BibitemShut {NoStop}%
\bibitem [{\citenamefont {Harris}\ and\ \citenamefont
  {Jones}(1974)}]{HarJon-JPF-74}%
  \BibitemOpen
  \bibfield  {author} {\bibinfo {author} {\bibfnamefont {J.}~\bibnamefont
  {Harris}}\ and\ \bibinfo {author} {\bibfnamefont {R.}~\bibnamefont {Jones}},\
  }\href@noop {} {\bibfield  {journal} {\bibinfo  {journal} {Journal of Physics
  F: Metal Physics}\ }\textbf {\bibinfo {volume} {4}},\ \bibinfo {pages} {1170}
  (\bibinfo {year} {1974})}\BibitemShut {NoStop}%
\bibitem [{\citenamefont {Gunnarsson}\ and\ \citenamefont
  {Lundqvist}(1976)}]{GunLun-PRB-76}%
  \BibitemOpen
  \bibfield  {author} {\bibinfo {author} {\bibfnamefont {O.}~\bibnamefont
  {Gunnarsson}}\ and\ \bibinfo {author} {\bibfnamefont {B.~I.}\ \bibnamefont
  {Lundqvist}},\ }\href@noop {} {\bibfield  {journal} {\bibinfo  {journal}
  {Physical Review B}\ }\textbf {\bibinfo {volume} {13}},\ \bibinfo {pages}
  {4274} (\bibinfo {year} {1976})}\BibitemShut {NoStop}%
\bibitem [{\citenamefont {Langreth}\ and\ \citenamefont
  {Perdew}(1975)}]{LanPer-SSC-75}%
  \BibitemOpen
  \bibfield  {author} {\bibinfo {author} {\bibfnamefont {D.~C.}\ \bibnamefont
  {Langreth}}\ and\ \bibinfo {author} {\bibfnamefont {J.~P.}\ \bibnamefont
  {Perdew}},\ }\href@noop {} {\bibfield  {journal} {\bibinfo  {journal} {Solid.
  State Commun.}\ }\textbf {\bibinfo {volume} {17}},\ \bibinfo {pages} {1425}
  (\bibinfo {year} {1975})}\BibitemShut {NoStop}%
\bibitem [{\citenamefont {Langreth}(1984)}]{Lan-PRL-84}%
  \BibitemOpen
  \bibfield  {author} {\bibinfo {author} {\bibfnamefont {D.~C.}\ \bibnamefont
  {Langreth}},\ }\href@noop {} {\bibfield  {journal} {\bibinfo  {journal}
  {Physical Review Letters}\ }\textbf {\bibinfo {volume} {52}},\ \bibinfo
  {pages} {2317} (\bibinfo {year} {1984})}\BibitemShut {NoStop}%
\bibitem [{\citenamefont {G{\"o}rling}\ and\ \citenamefont
  {Levy}(1993)}]{GorLev-PRB-93}%
  \BibitemOpen
  \bibfield  {author} {\bibinfo {author} {\bibfnamefont {A.}~\bibnamefont
  {G{\"o}rling}}\ and\ \bibinfo {author} {\bibfnamefont {M.}~\bibnamefont
  {Levy}},\ }\href@noop {} {\bibfield  {journal} {\bibinfo  {journal} {Physical
  Review B}\ }\textbf {\bibinfo {volume} {47}},\ \bibinfo {pages} {13105}
  (\bibinfo {year} {1993})}\BibitemShut {NoStop}%
\bibitem [{\citenamefont {G\"{o}rling}\ and\ \citenamefont
  {Levy}(1994)}]{GorLev-PRA-94}%
  \BibitemOpen
  \bibfield  {author} {\bibinfo {author} {\bibfnamefont {A.}~\bibnamefont
  {G\"{o}rling}}\ and\ \bibinfo {author} {\bibfnamefont {M.}~\bibnamefont
  {Levy}},\ }\href@noop {} {\bibfield  {journal} {\bibinfo  {journal} {Physical
  Review A}\ }\textbf {\bibinfo {volume} {50}},\ \bibinfo {pages} {196}
  (\bibinfo {year} {1994})}\BibitemShut {NoStop}%
\bibitem [{\citenamefont {Seidl}(1999)}]{Sei-PRA-99}%
  \BibitemOpen
  \bibfield  {author} {\bibinfo {author} {\bibfnamefont {M.}~\bibnamefont
  {Seidl}},\ }\href {\doibase 10.1103/PhysRevA.60.4387} {\bibfield  {journal}
  {\bibinfo  {journal} {Physical Review A}\ }\textbf {\bibinfo {volume} {60}},\
  \bibinfo {pages} {4387} (\bibinfo {year} {1999})}\BibitemShut {NoStop}%
\bibitem [{\citenamefont {Gori-Giorgi}\ \emph {et~al.}(2009)\citenamefont
  {Gori-Giorgi}, \citenamefont {Vignale},\ and\ \citenamefont
  {Seidl}}]{GorVigSei-JCTC-09}%
  \BibitemOpen
  \bibfield  {author} {\bibinfo {author} {\bibfnamefont {P.}~\bibnamefont
  {Gori-Giorgi}}, \bibinfo {author} {\bibfnamefont {G.}~\bibnamefont
  {Vignale}}, \ and\ \bibinfo {author} {\bibfnamefont {M.}~\bibnamefont
  {Seidl}},\ }\href {\doibase 10.1021/ct8005248} {\bibfield  {journal}
  {\bibinfo  {journal} {J. Chem. Theory Comput.}\ }\textbf {\bibinfo {volume}
  {5}},\ \bibinfo {pages} {743} (\bibinfo {year} {2009})}\BibitemShut {NoStop}%
\bibitem [{\citenamefont {Seidl}\ \emph {et~al.}(2007)\citenamefont {Seidl},
  \citenamefont {Gori-Giorgi},\ and\ \citenamefont {Savin}}]{SeiGorSav-PRA-07}%
  \BibitemOpen
  \bibfield  {author} {\bibinfo {author} {\bibfnamefont {M.}~\bibnamefont
  {Seidl}}, \bibinfo {author} {\bibfnamefont {P.}~\bibnamefont {Gori-Giorgi}},
  \ and\ \bibinfo {author} {\bibfnamefont {A.}~\bibnamefont {Savin}},\ }\href
  {\doibase 10.1103/PhysRevA.75.042511} {\bibfield  {journal} {\bibinfo
  {journal} {Phys. Rev. A}\ }\textbf {\bibinfo {volume} {75}},\ \bibinfo
  {pages} {042511/12} (\bibinfo {year} {2007})}\BibitemShut {NoStop}%
\bibitem [{\citenamefont {Fabiano}\ \emph {et~al.}(2016)\citenamefont
  {Fabiano}, \citenamefont {Gori-Giorgi}, \citenamefont {Seidl},\ and\
  \citenamefont {Della~Sala}}]{FabGorSeiDel-JCTC-16}%
  \BibitemOpen
  \bibfield  {author} {\bibinfo {author} {\bibfnamefont {E.}~\bibnamefont
  {Fabiano}}, \bibinfo {author} {\bibfnamefont {P.}~\bibnamefont
  {Gori-Giorgi}}, \bibinfo {author} {\bibfnamefont {M.}~\bibnamefont {Seidl}},
  \ and\ \bibinfo {author} {\bibfnamefont {F.}~\bibnamefont {Della~Sala}},\
  }\href@noop {} {\bibfield  {journal} {\bibinfo  {journal} {J. Chem. Theory.
  Comput.}\ }\textbf {\bibinfo {volume} {12}},\ \bibinfo {pages} {4885}
  (\bibinfo {year} {2016})}\BibitemShut {NoStop}%
\bibitem [{\citenamefont {Vuckovic}\ \emph {et~al.}(2018)\citenamefont
  {Vuckovic}, \citenamefont {Gori-Giorgi}, \citenamefont {Della~Sala},\ and\
  \citenamefont {Fabiano}}]{VucGorDelFab-JPCL-18}%
  \BibitemOpen
  \bibfield  {author} {\bibinfo {author} {\bibfnamefont {S.}~\bibnamefont
  {Vuckovic}}, \bibinfo {author} {\bibfnamefont {P.}~\bibnamefont
  {Gori-Giorgi}}, \bibinfo {author} {\bibfnamefont {F.}~\bibnamefont
  {Della~Sala}}, \ and\ \bibinfo {author} {\bibfnamefont {E.}~\bibnamefont
  {Fabiano}},\ }\href {\doibase 10.1021/acs.jpclett.8b01054} {\bibfield
  {journal} {\bibinfo  {journal} {J. Phys. Chem. Lett.}\ }\textbf {\bibinfo
  {volume} {9}},\ \bibinfo {pages} {3137} (\bibinfo {year} {2018})}\BibitemShut
  {NoStop}%
\bibitem [{\citenamefont {Giarrusso}\ \emph {et~al.}(2018)\citenamefont
  {Giarrusso}, \citenamefont {Gori-Giorgi}, \citenamefont {Della~Sala},\ and\
  \citenamefont {Fabiano}}]{GiaGorDelFab-JCP-18}%
  \BibitemOpen
  \bibfield  {author} {\bibinfo {author} {\bibfnamefont {S.}~\bibnamefont
  {Giarrusso}}, \bibinfo {author} {\bibfnamefont {P.}~\bibnamefont
  {Gori-Giorgi}}, \bibinfo {author} {\bibfnamefont {F.}~\bibnamefont
  {Della~Sala}}, \ and\ \bibinfo {author} {\bibfnamefont {E.}~\bibnamefont
  {Fabiano}},\ }\href {\doibase 10.1063/1.5022669} {\bibfield  {journal}
  {\bibinfo  {journal} {J. Chem. Phys.}\ }\textbf {\bibinfo {volume} {148}},\
  \bibinfo {pages} {134106} (\bibinfo {year} {2018})}\BibitemShut {NoStop}%
\bibitem [{\citenamefont {Daas}\ \emph {et~al.}(2021)\citenamefont {Daas},
  \citenamefont {Fabiano}, \citenamefont {Della~Sala}, \citenamefont
  {Gori-Giorgi},\ and\ \citenamefont {Vuckovic}}]{DaaFabSalGorVuc-JPCL-21}%
  \BibitemOpen
  \bibfield  {author} {\bibinfo {author} {\bibfnamefont {T.~J.}\ \bibnamefont
  {Daas}}, \bibinfo {author} {\bibfnamefont {E.}~\bibnamefont {Fabiano}},
  \bibinfo {author} {\bibfnamefont {F.}~\bibnamefont {Della~Sala}}, \bibinfo
  {author} {\bibfnamefont {P.}~\bibnamefont {Gori-Giorgi}}, \ and\ \bibinfo
  {author} {\bibfnamefont {S.}~\bibnamefont {Vuckovic}},\ }\href@noop {}
  {\bibfield  {journal} {\bibinfo  {journal} {J. Phys. Chem. Lett.}\ }\textbf
  {\bibinfo {volume} {12}},\ \bibinfo {pages} {4867} (\bibinfo {year}
  {2021})}\BibitemShut {NoStop}%
\bibitem [{\citenamefont {Pernal}(2018)}]{Per-IJQC-18}%
  \BibitemOpen
  \bibfield  {author} {\bibinfo {author} {\bibfnamefont {K.}~\bibnamefont
  {Pernal}},\ }\href@noop {} {\bibfield  {journal} {\bibinfo  {journal}
  {International Journal of Quantum Chemistry}\ }\textbf {\bibinfo {volume}
  {118}},\ \bibinfo {pages} {e25462} (\bibinfo {year} {2018})}\BibitemShut
  {NoStop}%
\bibitem [{\citenamefont {Seidl}\ \emph {et~al.}(2018)\citenamefont {Seidl},
  \citenamefont {Giarrusso}, \citenamefont {Vuckovic}, \citenamefont
  {Fabiano},\ and\ \citenamefont {Gori-Giorgi}}]{SeiGiaVucFabGor-JCP-18}%
  \BibitemOpen
  \bibfield  {author} {\bibinfo {author} {\bibfnamefont {M.}~\bibnamefont
  {Seidl}}, \bibinfo {author} {\bibfnamefont {S.}~\bibnamefont {Giarrusso}},
  \bibinfo {author} {\bibfnamefont {S.}~\bibnamefont {Vuckovic}}, \bibinfo
  {author} {\bibfnamefont {E.}~\bibnamefont {Fabiano}}, \ and\ \bibinfo
  {author} {\bibfnamefont {P.}~\bibnamefont {Gori-Giorgi}},\ }\href@noop {}
  {\bibfield  {journal} {\bibinfo  {journal} {J. Chem. Phys.}\ }\textbf
  {\bibinfo {volume} {149}},\ \bibinfo {pages} {241101} (\bibinfo {year}
  {2018})}\BibitemShut {NoStop}%
\bibitem [{\citenamefont {Helgaker}\ \emph {et~al.}(2014)\citenamefont
  {Helgaker}, \citenamefont {Jorgensen},\ and\ \citenamefont
  {Olsen}}]{purplebible}%
  \BibitemOpen
  \bibfield  {author} {\bibinfo {author} {\bibfnamefont {T.}~\bibnamefont
  {Helgaker}}, \bibinfo {author} {\bibfnamefont {P.}~\bibnamefont {Jorgensen}},
  \ and\ \bibinfo {author} {\bibfnamefont {J.}~\bibnamefont {Olsen}},\
  }\href@noop {} {\emph {\bibinfo {title} {Molecular electronic-structure
  theory}}}\ (\bibinfo  {publisher} {John Wiley \& Sons},\ \bibinfo {year}
  {2014})\BibitemShut {NoStop}%
\bibitem [{\citenamefont {Liu}\ and\ \citenamefont
  {Burke}(2009)}]{LiuBur-PRA-09}%
  \BibitemOpen
  \bibfield  {author} {\bibinfo {author} {\bibfnamefont {Z.-F.}\ \bibnamefont
  {Liu}}\ and\ \bibinfo {author} {\bibfnamefont {K.}~\bibnamefont {Burke}},\
  }\href@noop {} {\bibfield  {journal} {\bibinfo  {journal} {Physical Review
  A}\ }\textbf {\bibinfo {volume} {79}},\ \bibinfo {pages} {064503} (\bibinfo
  {year} {2009})}\BibitemShut {NoStop}%
\bibitem [{\citenamefont {Seidl}\ \emph {et~al.}(1999)\citenamefont {Seidl},
  \citenamefont {Perdew},\ and\ \citenamefont {Levy}}]{SeiPerLev-PRA-99}%
  \BibitemOpen
  \bibfield  {author} {\bibinfo {author} {\bibfnamefont {M.}~\bibnamefont
  {Seidl}}, \bibinfo {author} {\bibfnamefont {J.~P.}\ \bibnamefont {Perdew}}, \
  and\ \bibinfo {author} {\bibfnamefont {M.}~\bibnamefont {Levy}},\ }\href
  {\doibase 10.1103/PhysRevA.59.51} {\bibfield  {journal} {\bibinfo  {journal}
  {Physical Review A}\ }\textbf {\bibinfo {volume} {59}},\ \bibinfo {pages}
  {51} (\bibinfo {year} {1999})}\BibitemShut {NoStop}%
\bibitem [{\citenamefont {M{\o}ller}\ and\ \citenamefont
  {Plesset}(1934)}]{MolPle-PR-34}%
  \BibitemOpen
  \bibfield  {author} {\bibinfo {author} {\bibfnamefont {C.}~\bibnamefont
  {M{\o}ller}}\ and\ \bibinfo {author} {\bibfnamefont {M.~S.}\ \bibnamefont
  {Plesset}},\ }\href {\doibase 10.1103/physrev.46.618} {\bibfield  {journal}
  {\bibinfo  {journal} {Physical Review}\ }\textbf {\bibinfo {volume} {46}},\
  \bibinfo {pages} {618} (\bibinfo {year} {1934})}\BibitemShut {NoStop}%
\bibitem [{\citenamefont {Perdew}\ \emph {et~al.}(1996)\citenamefont {Perdew},
  \citenamefont {Burke},\ and\ \citenamefont {Ernzerhof}}]{PerBurErn-PRL-96}%
  \BibitemOpen
  \bibfield  {author} {\bibinfo {author} {\bibfnamefont {J.~P.}\ \bibnamefont
  {Perdew}}, \bibinfo {author} {\bibfnamefont {K.}~\bibnamefont {Burke}}, \
  and\ \bibinfo {author} {\bibfnamefont {M.}~\bibnamefont {Ernzerhof}},\
  }\href@noop {} {\bibfield  {journal} {\bibinfo  {journal} {Physical Review
  Letters}\ }\textbf {\bibinfo {volume} {77}},\ \bibinfo {pages} {3865}
  (\bibinfo {year} {1996})}\BibitemShut {NoStop}%
\bibitem [{\citenamefont {Adamo}\ and\ \citenamefont {Barone}(1999)}]{pbe0_2}%
  \BibitemOpen
  \bibfield  {author} {\bibinfo {author} {\bibfnamefont {C.}~\bibnamefont
  {Adamo}}\ and\ \bibinfo {author} {\bibfnamefont {V.}~\bibnamefont {Barone}},\
  }\href {\doibase 10.1063/1.478522} {\bibfield  {journal} {\bibinfo  {journal}
  {J. Chem. Phys.}\ }\textbf {\bibinfo {volume} {110}},\ \bibinfo {pages}
  {6158} (\bibinfo {year} {1999})}\BibitemShut {NoStop}%
\bibitem [{\citenamefont {Carrascal}\ \emph {et~al.}(2016)\citenamefont
  {Carrascal}, \citenamefont {Ferrer}, \citenamefont {Smith},\ and\
  \citenamefont {Burke}}]{CarFerSmiBur-JPCM-16}%
  \BibitemOpen
  \bibfield  {author} {\bibinfo {author} {\bibfnamefont {D.}~\bibnamefont
  {Carrascal}}, \bibinfo {author} {\bibfnamefont {J.}~\bibnamefont {Ferrer}},
  \bibinfo {author} {\bibfnamefont {J.}~\bibnamefont {Smith}}, \ and\ \bibinfo
  {author} {\bibfnamefont {K.}~\bibnamefont {Burke}},\ }\href@noop {}
  {\bibfield  {journal} {\bibinfo  {journal} {Journal of Physics. Condensed
  Matter}\ }\textbf {\bibinfo {volume} {29}} (\bibinfo {year}
  {2016})}\BibitemShut {NoStop}%
\bibitem [{\citenamefont {Seidl}\ \emph {et~al.}(2000)\citenamefont {Seidl},
  \citenamefont {Perdew},\ and\ \citenamefont {Kurth}}]{SeiPerKur-PRA-00}%
  \BibitemOpen
  \bibfield  {author} {\bibinfo {author} {\bibfnamefont {M.}~\bibnamefont
  {Seidl}}, \bibinfo {author} {\bibfnamefont {J.~P.}\ \bibnamefont {Perdew}}, \
  and\ \bibinfo {author} {\bibfnamefont {S.}~\bibnamefont {Kurth}},\
  }\href@noop {} {\bibfield  {journal} {\bibinfo  {journal} {Physical Review
  A}\ }\textbf {\bibinfo {volume} {{62}}},\ \bibinfo {pages} {012502} (\bibinfo
  {year} {2000})}\BibitemShut {NoStop}%
\bibitem [{\citenamefont {Daas}\ \emph {et~al.}(2022)\citenamefont {Daas},
  \citenamefont {Kooi}, \citenamefont {Grooteman}, \citenamefont {Seidl},\ and\
  \citenamefont {Gori-Giorgi}}]{DaaKooGroSeiGor-JCTC-22}%
  \BibitemOpen
  \bibfield  {author} {\bibinfo {author} {\bibfnamefont {T.~J.}\ \bibnamefont
  {Daas}}, \bibinfo {author} {\bibfnamefont {D.~P.}\ \bibnamefont {Kooi}},
  \bibinfo {author} {\bibfnamefont {A.~J.}\ \bibnamefont {Grooteman}}, \bibinfo
  {author} {\bibfnamefont {M.}~\bibnamefont {Seidl}}, \ and\ \bibinfo {author}
  {\bibfnamefont {P.}~\bibnamefont {Gori-Giorgi}},\ }\href@noop {} {\bibfield
  {journal} {\bibinfo  {journal} {J. Chem. Theory Comput.}\ }\textbf {\bibinfo
  {volume} {18}},\ \bibinfo {pages} {1584} (\bibinfo {year}
  {2022})}\BibitemShut {NoStop}%
\bibitem [{\citenamefont {{\'S}miga}\ \emph {et~al.}(2022)\citenamefont
  {{\'S}miga}, \citenamefont {Della~Sala}, \citenamefont {Gori-Giorgi},\ and\
  \citenamefont {Fabiano}}]{SmiSalGorFab-arxiv-21}%
  \BibitemOpen
  \bibfield  {author} {\bibinfo {author} {\bibfnamefont {S.}~\bibnamefont
  {{\'S}miga}}, \bibinfo {author} {\bibfnamefont {F.}~\bibnamefont
  {Della~Sala}}, \bibinfo {author} {\bibfnamefont {P.}~\bibnamefont
  {Gori-Giorgi}}, \ and\ \bibinfo {author} {\bibfnamefont {E.}~\bibnamefont
  {Fabiano}},\ }\href@noop {} {\bibfield  {journal} {\bibinfo  {journal} {arXiv
  preprint arXiv:2202.11531}\ } (\bibinfo {year} {2022})}\BibitemShut {NoStop}%
\end{thebibliography}%


\end{document}